\documentclass[11pt]{article}
\usepackage{amsmath,amssymb,mcite,citesort}
\usepackage{rotating}
\usepackage{caption}
\renewcommand{\d}{\textup{d}}
\newcommand{\e}{\textup{e}}

\begin{document}

\begin{titlepage}

\hfill AEI-2013-217

\vspace{2.5cm}
\begin{center}

{{\LARGE  \bf 	Duality completion of higher derivative corrections}} \\

\vskip 1.75cm {Hadi Godazgar
and Mahdi Godazgar}
\\
{\vskip 0.5cm
Max-Planck-Institut f\"{u}r Gravitationsphysik, \\
Albert-Einstein-Institut,\\
Am M\"{u}hlenberg 1, D-14476 Potsdam, Germany
}
{\vskip 0.35cm
hadi.godazgar@aei.mpg.de, mahdi.godazgar@aei.mpg.de}
\end{center}

\vskip 0.5cm

\begin{center}
\today
\end{center}

\noindent

\vskip 1.75cm

\begin{abstract}
\noindent We present a new method for completing higher derivative corrections for theories that exhibit duality symmetries under reduction.  This proposal is based on the observation that duality symmetry in the reduced theory highly constrains the form of the unreduced theory.  We apply this idea to closed bosonic string theory and complete the Riemann squared term to simply derive the known full tree-level effective action to order $\alpha'$.  
\end{abstract}

\end{titlepage}

\section{Introduction}

A crucial ingredient in understanding the role, or otherwise, of string theory in nature is to understand string theory beyond the low energy effective supergravity description. Of course, supergravity plays an important role in guiding this understanding, particularly via non-perturbative insights. However, it is ultimately corrections to the supergravity limit that makes string theory a desirable candidate for a theory of quantum gravity. These corrections can be viewed as coming from two distinct sources: $\alpha'$ corrections, which arise due to the finite length of the fundamental string, and quantum corrections in the string coupling $g_s.$ From a spacetime point of view, the $\alpha'$ expansion corresponds to a higher derivative expansion beyond classical two-derivative supergravity, while the $g_s$ expansion contributes at each given order in derivatives.

The ultimate goal is to find an effective action that incorporates all such corrections, including non-perturbative effects.  This is clearly a difficult problem.  A more modest starting point in this direction is to understand these corrections order by order.  The most direct approach is string perturbation theory. The novelty of string theory is that at each order in the quantum expansion there are an infinite number of terms in an expansion in $\alpha'$. A remarkable observation is that the equations of motion for the massless states found from the string S-matrix coincide perturbatively with the vanishing of the $\beta$-functional of the string sigma model, thereby ensuring conformal invariance. It is expected that this equivalence will hold to all orders. For string theories admitting a Green-Schwarz formulation, the $\kappa$-invariance of the action can also be used to determine higher derivative corrections. This is one of the ways in which string theory is different from field theories, where the 
effective action is solely determined by the S-matrix.

In light of the fact that the low energy effective action is constrained, or even determined, in a number of ways, it is reasonable to ask in what way the duality symmetry of the action under reduction constrains the higher derivative corrections.  The existence of duality symmetries is a ubiquitous feature of gravitational theories.  In the context of general relativity, it has been known for a long time that a rich, unexpected symmetry structure appears in the presence of Killing isometries; a fact that has been utilised extensively to generate new interesting solutions. In supergravity theories, the inclusion of form-fields and fermions results in a yet richer structure \cite{CJ, Julia}.  Viewing supergravity as a low energy effective description of string theory, a relation is made \cite{maharanaschwarz} between these duality symmetries and the duality symmetries, such as T-duality, emerging from string theory, leading to the conjecture \cite{mtheoryht} that discrete versions of the Cremmer-Julia 
exceptional symmetry groups are in fact duality symmetries of string theories that encompass T and S-dualities.  

Using duality symmetries to understand new aspects of the theory, such as constraining or determining higher derivative corrections, is not a new idea and indeed significant work has been done in this regard\footnote{A non-exhaustive list of references on duality and higher derivative corrections in the context of string/M-theory includes \cite{greengutperle,Kiritsis:1997em,Obers:1999es,Berkovits:2004px,Damour:2005zb,Green:2005ba,Lambert:2006ny,Basu:2007ru,Garousi:2009dj,Green:2010wi,Pioline:2010kb,Green:2010sp,Green:2010kv,Becker:2010ij,Gubay:2011jk,McOrist:2012yc,Garousi:2012yr,Fleig:2012zc,Liu:2013dna}. For work on duality and higher derivative corrections in the context of four-dimensional gravity see \cite{Michel:2007vh,Colonnello:2007qy,Biswas:2011ar}.}.  In this paper, however, we present a new method for completing higher derivative corrections given the existence of duality symmetries under reduction.  In particular, we show how duality in the \emph{reduced} theory can be used to provide a completion of a higher derivative term in the \emph{unreduced} theory. Our approach is based on the observation \cite{MV91a, MV91b, GMV, maharanaschwarz} that string dualities imply that the reduced 
low energy effective action can be written in a duality manifest way in terms of a duality group element \emph{\`a la} Cremmer-Julia. This observation can also be extended to the higher derivative corrections. Specifically, Meissner \cite{meissner} showed that the $\alpha'$ corrections to the closed bosonic string low energy effective action when reduced to one dimension can be expressed solely in terms of a duality invariant dilaton field and an O$(d,d)$ group element. The new perspective that we have in this paper is that the insistence that the reduced theory, in particular the scalars, be written in terms of a group-theoretic duality element highly constrains, to the point of uniqueness, the unreduced theory.  Thus, rather than finding the duality group element by reducing the known action and explicitly demonstrating the duality invariance of the reduced theory, we show how given a duality symmetry, the unreduced action can be derived.  The method presented in this paper is particularly important given 
the fact that in many cases only a part of a higher derivative correction is known.  In general, the most well-understood terms are purely gravitational terms, which using our approach can be completed to find couplings to matter fields. The higher derivative corrections including matter couplings are important in a number of areas, including string phenomenology and cosmology and string theoretic studies of black hole entropy.  

In section \ref{sec:dualcomp}, we present the arguments for the duality completion of known terms in a theory given the existence of a duality symmetry. Central to this method is the observation that the scalars in the reduced theory parametrise the duality coset. These scalars are related to the internal components of higher dimensional fields.  We argue that general covariance provides a precise relationship between the action for the internal components of the fields and the full higher-dimensional (unreduced) action. Therefore, by completing the scalars from the reduction of a known term in the reduced theory, we can uplift to find the completion in the unreduced theory. 

To demonstrate the utility of this method, we apply our reasoning to find the T-duality completion of the Riemann squared correction to the low energy effective action of closed bosonic string theory and rederive the full tree-level correction at first order in $\alpha'$. We give a detailed derivation of our result in order to emphasise the simplicity of the method and the uniqueness of the completion. Furthermore, we would like to highlight the constructive nature of this method and provide the necessary framework for applying it to other theories.

First, in section \ref{secOddcoset}, we use group theory to find the duality group element that the scalars of the reduced theory parametrise.  Then, in section \ref{secbos}, we begin with the Riemann squared term and find the action for the scalars coming from the reduction of this term.  We rewrite this action using the duality group element with all other fields turned off.  Then, we turn on the other fields and calculate the scalar action in the reduced theory.  Finally, we uplift this new scalar action to the full theory.  Remarkably, we find that the uplift is unique and, up to field redefinitions, coincides precisely with the complete tree-level closed bosonic effective action to order $\alpha'$ \cite{grosssloan, metsaevtseytlin, HThigh, jackjones, osborn}.  Of significant importance is the simplicity with which these terms are found using this approach in stark contrast to previous derivations in the literature. 

We comment on the application of this work to other string theory corrections, inclusion of fermions and its relation to generalised geometry and double field theory in the discussion section at the end of the paper. 

Our index conventions throughout the paper are as follows: $D$-dimensional indices are denoted $\mu, \nu, \rho \ldots$; $d$-dimensional (internal) indices are denoted by $a, b, c,\ldots$ and $(D-d)$-dimensional (reduced) indices are denoted by $i, j, k, \ldots$.

\section{Duality completion of a gravitational sector} \label{sec:dualcomp}

In this section, we present an argument as to how one can use consistency with a duality symmetry that appears upon a dimensional reduction of a gravitational theory to determine the coupling of the gravitational sector to the matter content in the full theory. The argument presented here applies more generally for any sector of the theory and need not be confined to a completion of the gravitational sector.  However, given that it is generally the gravitational sector that is most well-understood in any given theory we shall confine our attention to this sector for clarity.

Consider a $D$-dimensional gravitational theory with some matter content.  In addition, assume that upon a reduction of the theory to $(D-d)$ dimensions hidden symmetries appear. That is, the reduced theory possesses a symmetry, given by a coset $G/H$, that is larger than that which one would naively expect.  The precise way in which the coset $G/H$ controls the dynamics of the reduced theory is only important here at the level of the scalars.  In particular, what we require here is the following:
\begin{quote}{\emph{Upon a dimensional reduction of the $D$-dimensional gravitational theory to $(D-d)$ dimensions, the scalars of the reduced theory parameterise a coset $G/H$.  Moreover, the scalar sector of the theory can be written, solely, in terms of the metric of the reduced theory, a coset element of $G/H$ and any scalar invariant under the action of $G/H$.}}\end{quote}
In some cases, an appropriate dualisation of fields needs to be performed in order to find the scalar sector of the reduced theory.  For the purposes of this argument, we neglect this possibility and only consider the case where the scalar sector of the reduced theory manifests itself naturally upon dimensional reduction and dualisation of fields need not be carried out.  A simple modification of the argument presented here can be applied to the former case as will hopefully become apparent.  However, for simplicity, we disregard this possibility here.

The claim, which we shall clarify below, is that any (non-gauge invariant) term that must appear in the $D$-dimensional action for consistency with the hidden symmetry observed upon a reduction of the theory can be completed to a unique gauge invariant term that is consistent with the appearance of the hidden symmetry in lower dimensions.  This is a non-trivial statement for two reasons.  It is certainly not necessarily true that any non-gauge invariant term can be completed to a gauge invariant term in the sense described above; nor should it be obvious to the reader why any such term must be unique.

Let us begin by splitting the lagrangian of the $D$-dimensional theory in the following way:
\begin{equation}
 L =\mathcal{R}_S(\tilde{\partial}, f) + \mathcal{R}_R(\tilde{\partial}, f, F) + \mathcal{G}(\hat{\partial}, f)
 + \mathcal{C}(\tilde{\partial}, \hat{\partial}, f, F),
 \label{splitL}
\end{equation}
where $\tilde{\partial}$ denotes partial differentiation with respect to a $(D-d)$-dimensional coordinate, i.e. the coordinates of what would be the reduced theory if dimensional reduction were to be carried out; $\hat{\partial}$ denotes partial differentiation with respect to the complementary coordinates; $f$ denotes all fields with only $d$-dimensional (internal space) indices\footnote{We refer to these as internal space indices in view of the terminology that would be used in the reduced theory.} and $F$ denotes the remaining fields.  These could have mixed indices or have only $(D-d)$-dimensional indices.  In $\mathcal{R}_S(\tilde{\partial}, f)$, $F$ type fields are required in order to contract with the partial derivatives and form a scalar.  However, in the notation, the emphasis is on those fields on which the derivatives act.  Note that the decomposition above changes depending on whether one chooses to perform integration by parts on some terms.  Thus, there is an integration by parts ambiguity in the decomposition described above.  However, at the level of the scalars, which is what we are interested in here, this will not make a difference.  This is because, for the scalar sector, the integration by parts in the full theory is reflected in integration by parts in the reduced theory.

For concreteness, consider the following term
\begin{equation}
 g^{\mu \nu} \partial_{\mu} g^{\rho \sigma} \partial_{\nu} g_{\rho \sigma}
\end{equation}
that appears in the Ricci scalar and is thus present in the lagrangian of any gravitational theory based on Einstein's theory.
For this term,
\begin{align}
 \mathcal{R}_S(\tilde{\partial}, f) &= g^{i j} \partial_{i} g^{a b} \partial_{j} g_{a b}, \\
 \mathcal{R}_R(\tilde{\partial}, f, F) &= g^{i j} \partial_{i} g^{k l} \partial_{j} g_{k l} + 2 g^{i j} \partial_{i} g^{a k} \partial_{j} g_{a k},  \\
 \mathcal{G}(\hat{\partial}, f) &= g^{a b} \partial_{a} g^{c d} \partial_{b} g_{c d}, \\
 \mathcal{C}(\tilde{\partial}, \hat{\partial}, f, F) &= g^{a b} (\partial_{a} g^{i j} \partial_b g_{i j} 
 + 2 \partial_{a} g^{i c} \partial_b g_{i c}) + 2 g^{i a} (\partial_{i} g^{c d} \partial_a g_{c d} 
 + 2 \partial_{i} g^{j b} \partial_a g_{j b} + \partial_{i} g^{k l} \partial_a g_{k l}).
\end{align}

Together, $\mathcal{R}_S$ and $\mathcal{R}_R$ correspond to the reduced theory, while terms in $\mathcal{G}$ correspond to the sector of the theory for which a generalised geometric formulation may be possible. It is clear that such a splitting of the lagrangian can always be done\footnote{Of course, it is not possible to decompose the measure in such a way as to split the action into a piece corresponding to the reduced theory; a piece corresponding to the generalised geometry and the rest of the terms.  Thus, we should really be working at the level of the equations of motion rather than the action.  However, this is equivalent to splitting the lagrangian, as described, and assigning a particular measure to each term of interest.}.  The terms that interest us here are those contained in $\mathcal{R}_S$.  To reiterate, these are terms for which partial differentiation is taken only along $(D-d)$-dimensional coordinates and all fields on which the derivatives act appear only with internal space indices.  
From the perspective of the reduced theory, these terms are in
the scalar sector of the theory\footnote{If dualisation of fields were required in the reduced theory to determine the scalar sector, then these terms would only form a part of the scalar sector.  The other terms that would contribute to the scalar sector would then come from $\mathcal{R}_R$. The precise nature of these terms would obviously depend on the precise matter content and the value of $d$.}.  It is this observation that allows us to argue the duality completion of any gravitational term in the lagrangian.

Consider a lagrangian with a gravitational term. Dimensionally reduce the action to $(D-d)$ dimensions and consider the scalar sector of the reduced theory.  By assumption, the scalar sector can be rewritten in terms of a coset element of $G/H$ and any scalar invariant under the coset, so that
\begin{equation}
 \tilde{\mathcal{R}}_S(\tilde{g}, f) = \tilde{\mathcal{R}}_S(\tilde{g}, \mathcal{V}_{G/H}, \{\Phi\})\Big|_{\text{matter fields}=0}
\end{equation}
where $\tilde{g}$ is the metric and $f$ are the scalars of the reduced theory; $\mathcal{V}_{G/H}$ denotes a coset element of $G/H$ and $\{\Phi\}$ denotes the set of scalars that are formed from $f$ and are invariant under the action of the coset.  Thus, $\tilde{\mathcal{R}}_S(\tilde{g}, \mathcal{V}_{G/H}, \{\Phi\})$ gives the full coupling of the scalars corresponding to the matter fields in the reduced theory, given a particular gravitational term. 

Now, the question is how to extend this observation to the unreduced theory.  This is where the splitting of the lagrangian of the $D$-dimensional theory, described above, becomes useful.  The terms in $\tilde{\mathcal{R}}_S(\tilde{g}, f)$ are in one-to-one correspondence with the terms in $\mathcal{R}_S(\tilde{\partial}, f)$, as emphasised earlier.  That is, the terms in $\mathcal{R}_S(\tilde{\partial}, f)$ reduce trivially to terms in $\tilde{\mathcal{R}}_S(\tilde{g}, f)$.  As such, the structure of the terms in each are identical\footnote{The only difference is that when reduction is carried out, some fields are redefined so that they have the right transformation properties in the reduced theory.  For example, $g^{ij}$ is not the inverse of $g_{ij}$.  Whereas $\tilde{g}^{ij}$ is indeed the inverse of $\tilde{g}_{ij}$, as one would expect.}.

If the $D$-dimensional theory is to be consistent with the appearance of duality symmetry in the reduced theory it must contain terms that reduce to
$$
\tilde{\mathcal{R}}_S(\tilde{g}, \mathcal{V}_{G/H}, \{\Phi\}),
$$
which controls the coupling of the matter to the gravitational field.  From the argument set out above, we conclude that $\mathcal{R}_S$ must contain terms of the same structure as those contained in $\tilde{\mathcal{R}}_S(\tilde{g}, \mathcal{V}_{G/H}, \{\Phi\})$, once it is expanded in terms of the canonical fields in the theory.  Inspecting equation \eqref{splitL}, it is straightforward to determine what kind of terms must appear in the unsplit lagrangian in order to give $\mathcal{R}_S$ once splitting takes place. For example, given $\mathcal{R}_S$ of the form
$$
g^{i j} \partial_{i} g^{a b} \partial_{j} g_{a b} + g^{i j} g^{a b} \partial_{i} A_{a} \partial_{j} A_{b},
$$
where $A$ is a 1-form, the $D$-dimensional lagrangian must contain terms of the form 
$$
g^{\mu \nu} \partial_{\mu} g^{\rho \sigma} \partial_{\nu} g_{\rho \sigma} + g^{\mu \nu} g^{\rho \sigma} \partial_{\mu} A_{\rho} \partial_{\nu} A_{\sigma}.
$$
Of course, as is clear from the example above, such terms will certainly not be diffeomorphism nor gauge invariant.  However, if the $D$-dimensional action is to be diffeomorphism and gauge invariant, which is what one would expect, then one should be able to complete the non-gauge invariant terms into gauge invariant terms without spoiling the structure of $\mathcal{R}_S$. It is certainly not clear \textit{a priori} that this should be possible. In the example we consider, we find that this is indeed the case. Furthermore, we  find that any such gauge invariant completion leads to a unique $D$-dimensional lagrangian. While this task may seem daunting at the abstract level, in practice it is straightforward, as should be clear from the example below. The only subtlety in practical terms is finding the coset element $\mathcal{V}_{G/H}$.

Of course, depending on the value of $(D-d)$ this procedure will fail to (re)produce terms in the higher dimensional action that do not contribute to $\mathcal{R}_S$.  For example, if we consider a reduction of the theory to more than two dimensions, then it is not possible to find a two-derivative Chern-Simons term in the $D$-dimensional action. This means that one needs to reduce to a low enough dimension in order to be able to find all possible terms at a particular derivative order in the $D$-dimensional lagrangian.  For a two derivative theory, it suffices to reduce to two dimensions. In general, the duality coset of gravitational theories reduced to two dimensions is infinite-dimensional \cite{geroch, juliainf, nicolaie9}.  However, this will not be important for the purposes of this paper as we are principally concerned with completing higher derivative terms.

To clarify the abstract arguments presented above, consider the specific example of a gravitational theory with a $2$-form potential $B$ and a scalar $\phi$.  In addition, assume that upon a toroidal reduction on $T^{d}$, the scalars of the reduced theory parameterise the coset 
O$(d,d)/$O$(d)\times$O$(d)$ \cite{maharanaschwarz}.  Furthermore, assume that the gravitational sector of the theory is given by the following action
\begin{equation}
 S_{grav}= \int d^{D}x \sqrt{det(g_{\mu \nu})} e^{-2 \phi} R(g_{\mu \nu}),
 \label{EHdilAction}
\end{equation}
where $R(g_{\mu \nu})$ is the Ricci scalar of metric $g_{\mu \nu}$.  This theory should be familiar to the reader. It can be identified with the low energy limit of closed bosonic string theory for $D=26$; heterotic string theory with non-abelian gauge fields turned off for $D=10$ and the NSNS sector of type II string theories for $D=10$.

Define the T-duality invariant field \cite{buscher1, buscher2}\footnote{It should be obvious from the context whether $d$ refers to this new field or the dimension of the internal space.}
\begin{equation} \label{measinv}
 e^{-2 d} = \sqrt{\hat{g}} e^{-2 \phi}.
\end{equation}
Equivalently,
\begin{equation}
\text{log} \hat{g}= 4(\phi-d),
\end{equation}
where $\hat{g}$ is the internal metric.

Using the results of appendix \ref{dimred}, we find that in the reduction to $(D-d)$ dimensions,
\begin{align}
 \sqrt{- g} e^{-2 \phi} R(g_{\mu \nu}) &= \sqrt{- \tilde{g}} \sqrt{\hat{g}} e^{-2 \phi} \left(
  R(\tilde{g}) - \frac{1}{4} \hat{g}  \tilde{g}^{i k} \tilde{g}^{jl} \hat{g}_{ab} f_{ij}{}^{a} f_{kl}{}^{b}
 + \frac{1}{4} \tilde{g}^{ij} (\partial_{i} \hat{g}^{ab})( \partial_{j} \hat{g}_{ab} )  \notag \right.\\[5pt] 
& \hspace{55mm} \left. + \frac{1}{4} \tilde{g}^{ij} (\hat{g}^{ab} \partial_{i} \hat{g}_{ab}) (\hat{g}^{cd} \partial_{j} \hat{g}_{cd}) -2 \hat{g}^{-1} \partial_{i} \hat{g} \partial_{j} \phi\right),
 \label{redgrav}
\end{align}
where $g\equiv det(g_{\mu \nu})$ and $$f_{ij}{}^{a} = 2 \partial_{[i} b_{j]}{}^{a}. $$  
Assuming that $(D-d)>3$ so that the $d$ 1-forms $b_{i}{}^{a}$ cannot be dualised into scalars, the field content of the reduced theory includes a metric $\tilde{g}_{ij}$, $d$ 1-forms $b_{i}{}^{a}$ and $d(d+1)/2$ scalars coming from the internal metric $\hat{g}_{ab}$.  

The third, fourth and fifth terms on the right hand side of equation \eqref{redgrav} correspond to the scalar sector of the theory.  Given the arguments above one should be able to embed this sector into an expression written only in terms of a coset element of O$(d,d)/$O$(d)\times$O$(d)$ and the invariant measure $d$, defined in equation \eqref{measinv}. 

The coset element of O$(d,d)/$O$(d)\times$O$(d)$ is after all a group theoretic object and can derived without any assumptions regarding the matter couplings of the theory. As explained in section \eqref{secOddcoset}, it is more convenient to work with the duality group element 
\begin{equation} 
G = \mathcal{V}_{G/H}^{T} \mathcal{V}_{G/H},
\end{equation}
which is derived in that section.  Written in a canonical form such that it coincides with the O$(d,d)$ group element that is familiar from the literature \cite{tseytlinodd, duff, maharanaschwarz}, 
\begin{equation}
 G = 
\begin{pmatrix}
 \hat{g} - B \hat{g}^{-1} B & B \hat{g}^{-1} \\
- \hat{g}^{-1} B & \hat{g}^{-1}
\end{pmatrix}, \label{Odd:G}
\end{equation}
where all indices in the matrix, which have been suppressed, take $d$-dimensional internal space values, i.e. they are of the form $a, b, c, \ldots$.

Consider \cite{maharanaschwarz}
\begin{equation} \label{Oddcov2der}
 \sqrt{-\tilde{g}} e^{-2d} \left( \frac{1}{8} \tilde{g}^{ij} \text{Tr} (\partial_{i} G^{-1} \partial_{j} G) + 4 \tilde{g}^{ij} \partial_{i} d \partial_{j} d \right).
\end{equation}
Expanding out the above term using the definition of $G$, equation \eqref{Odd:G}, and $d$, equation \eqref{measinv}, we find the terms in the scalar sector of the reduced theory above as well as other terms:
\begin{align} \label{Gexpand}
\sqrt{-\tilde{g}} e^{-2d} & \left( \frac{1}{8} \tilde{g}^{ij} \text{Tr} (\partial_{i} G^{-1} \partial_{j} G) + 4 \tilde{g}^{ij} \partial_{i} d \partial_{j} d \right) \notag \\
 &= \sqrt{- \tilde{g}} \sqrt{\hat{g}} e^{-2 \phi} \tilde{g}^{ij} \left( \frac{1}{4}  \partial_{i} \hat{g}^{ab} \partial_{j} \hat{g}_{ab} \right. + \frac{1}{4} (\hat{g}^{ab} \partial_{i} \hat{g}_{ab}) (\hat{g}^{cd} \partial_{j} \hat{g}_{cd}) -2 \hat{g}^{-1} \partial_{i} \hat{g} \partial_{j} \phi \notag \\
&\hspace{70mm} - \left. \frac{1}{4} \hat{g}^{ac} \hat{g}^{bd} \partial_{i}B_{ab} \partial_{j}B_{cd}  + 4 \partial_{i} \phi \partial_{j} \phi \right). 
\end{align}
Of course, the coefficients in \eqref{Oddcov2der} were fixed so as to recover the first two terms in the scalar sector.  Thus, the first two terms on the right hand side of the equation above appear by construction.  The third term in the scalar sector also appears as one would expect. Hence, we find that in order for the reduced theory to be invariant under the coset O$(d,d)/$O$(d)\times$O$(d)$, as has been assumed, the terms in the second line on the right hand side of the equation above must, also, appear in the scalar sector of the reduced theory.  From this knowledge, we now hope to construct the coupling of the matter fields to gravity in the $D$-dimensional lagrangian.

Inspecting the extra terms on the right hand side of equation \eqref{Gexpand}, it is clear that these can only come from a $D$-dimensional action containing the following terms
\begin{equation} \label{nongaugeaction}
 \int d^{D}x \sqrt{g} e^{-2 \phi} \left( -\frac{1}{4} g^{\mu \nu} g^{\rho \sigma} g^{\tau \eta} \partial_{\mu}B_{\rho \tau} \partial_{\nu}B_{\sigma \eta} + 4 g^{\mu \nu} \partial_{\mu} \phi \partial_{\nu} \phi \right).
\end{equation}
A gauge invariant term that contains the first term must be, schematically, of the form
\begin{equation}
 g^{-1} g^{-1} g^{-1} H H,
\end{equation}
where $H=dB$.  Furthermore, using the antisymmetric nature of $H$ and the nature of the contractions in the first term in \eqref{nongaugeaction}, we conclude that the gauge invariant term must be of the form
\begin{equation}
g^{\mu \nu} g^{\rho \sigma} g^{\tau \eta} H_{\mu \rho \tau} H_{\nu \sigma \eta} =3 g^{\mu \nu} g^{\rho \sigma} g^{\tau \eta} (\partial_{\mu}B_{\rho \tau} \partial_{\nu}B_{\sigma \eta} - 2 \partial_{\rho}B_{\mu \tau} \partial_{\nu}B_{\sigma \eta}).
\end{equation}
The second term in \eqref{nongaugeaction} is already diffeomorphism invariant.  Putting all this together gives the full action, which should be familiar to the reader
\begin{equation}
 S= \int d^{D}x \sqrt{g} e^{-2 \phi} \left(R + 4 (\partial \phi)^2 - \frac{1}{12} H^2 \right),
 \label{closed}
\end{equation}
where $(\partial \phi)^2=g^{\mu \nu} \partial_{\mu} \phi \partial_{\nu} \phi$ and $H^2=g^{\mu \nu} g^{\rho \sigma} g^{\tau \eta} H_{\mu \rho \tau} H_{\nu \sigma \eta}$.

\section{Duality group element}

In this section, we construct the duality group element required for the duality completion of the Riemann squared correction in closed bosonic string theory.  The relevant duality group in this case is the T-duality group, which is given by the indefinite orthogonal group O$(d,d).$  In particular, upon reduction, the scalars parametrise the corresponding duality coset O$(d,d)/$O$(d)\times$O$(d).$   Worldsheet arguments \cite{buscher2} show that this symmetry is valid to all orders in $\alpha',$ and even perturbatively in $g_s$ \cite{rocver}. However, this has only been shown from the spacetime perspective for the first order correction in $\alpha'$ in the reduction to one-dimension \cite{meissner}\footnote{We would like to thank Axel Kleinschmidt for drawing our attention to this work.}, see also \cite{senodd}. 

The analysis of Meissner shows that in order to realise T-duality in the aforementioned theory it is necessary to carry out \emph{non-gauge invariant} field redefinitions of the metric and the gauge 3-form. These field redefinitions can equivalently be regarded as an $\alpha'$ correction to the coset element. Note that this is somewhat dissimilar to the situation encountered when higher derivative corrections are constructed. Namely, the $\alpha'$ correction to the metric is given by the Ricci tensor and can be removed by a \emph{gauge invariant} field redefinition. Hence the higher derivative corrections are given in terms of the original fields. Therefore, even though there is a stringy modification to geometry, this can be removed by a field redefinition. However, when considering the duality-manifest formulation of the higher derivative corrections in terms of the coset element this is no longer possible.  Evidence for this comes from considerations in string field theory where diffeomorphisms and gauge 
transformations of the NS-NS 2-form are modified by $\alpha'$ corrections. 

Further evidence for the fact that the duality coset in string theory must be corrected comes from generalised geometry and double field theory \cite{dft1, *dft2, *dft3, *genmetstring}. This geometry extends the tangent space to include $2$-forms which results in a unification of diffeomorphisms and gauge invariance. In this framework, while it is possible to construct analogues of the Ricci tensor and scalar, the Riemann tensor has proved to be elusive \cite{gendiff1, *gendiff2, *gendiff3, *CSW, *hohmzwiebach2, *bbmp, hohmzwiebach1}. In fact it can be shown that there is no concomitant of the O$(d,d)$ generalised metric (which is the square of the duality coset element), its derivative and $\eta,$ which is the bilinear form of O$(d,d),$ that transforms as a connection. Moreover, the only objects that are formed from the above and second derivatives of the generalised metric that transform covariantly are the Ricci tensor and scalar curvature already found in double field theory \cite{noncon} (see also \cite{
gendiff1, *gendiff2, *gendiff3, *CSW, *hohmzwiebach2, *bbmp, hohmzwiebach1}). Hence in this section, using group theory, we construct the duality coset element in O$(d,d)$ that is 
corrected by higher derivative terms.  Note that these corrections can be viewed as non-gauge invariant field redefinitions.

In practice, we use the square of the duality coset element
\begin{equation} \label{cosetsquare}
\mathcal{G} = \mathcal{V}^{T} \mathcal{V},
\end{equation}
which we call the duality group element.  Of course, when considering fermions, which transform under O$(d)\times$O$(d)$, it is crucial to use the coset element, $\mathcal{V}$.  However, since we are only considering bosonic fields here, it is more convenient to use $\mathcal{G}$ because it is invariant under local O$(d)\times$O$(d)$ transformations.

\subsection{O$(d,d)/$O$(d)\times$O$(d)$ group element}
\label{secOddcoset}

The duality group element for O$(d,d)$ is well known from the literature \cite{tseytlinodd, duff, maharanaschwarz}. It is given in terms of the metric $g$ and the 2-form NSNS field by 
\begin{equation}
\begin{pmatrix}
g_{ab} - B_{a e} g^{e f} B_{f b}  & B_{a e} g^{e d} \\
 - g^{c e} B_{e b} & g^{cd}
 \end{pmatrix}.
\label{oddgenmet}
\end{equation}
This is the central object in T-duality manifest formulations such as double field theory where it is called the generalised metric.  We rederive the form of the duality group element using the fact that it is given by a generic element of the O$(d,d)/$O$(d)\times$O$(d)$ coset. This will be a warm up for finding the form of the $\alpha'$ corrected duality group element.

Let
\begin{equation}
 \mathcal{G}= G + \alpha' H + \mathcal{O}(\alpha'{}^2)
\label{Gdef}
\end{equation} 
be an element of O$(d,d).$  In particular, from equation \eqref{cosetsquare}, $\mathcal{G}$ is a symmetric matrix. Therefore, $\mathcal{G}$ satisfies
\begin{gather} 
 \mathcal{G}^{T}= \mathcal{G}, \label{Gsym} \\
\mathcal{G}^{T} \eta \mathcal{G} = \eta.
\label{orthog}
\end{gather}
The block $2d \times 2d$ matrix $\eta$ is 
$$\eta = 
\begin{pmatrix}
0 & 1 \\
1 & 0
\end{pmatrix},
$$
where $1$ and $0$ in $\eta$ denote the $d$-dimensional identity and zero matrices, respectively.
Hence, equation \eqref{orthog} implies that $\mathcal{G}$ is an indefinite orthogonal matrix. 

From equation \eqref{Gdef} and \eqref{Gsym}, it is clear that both $G$ and $H$ are symmetric. Furthermore, substituting equation \eqref{Gdef} into \eqref{orthog} gives
\begin{equation}
 G \eta G + \alpha' \left( G \eta H + H \eta G \right) + \mathcal{O}(\alpha'{}^2) = \eta.
\end{equation}
 Hence, 
\begin{gather}
  G \eta G = \eta, \label{m0eqn} \\
 G \eta H + H \eta G = 0. \label{m1eqn}
\end{gather}
To show that $G$ is of the form \eqref{oddgenmet}, let
\begin{equation}
 G = 
\begin{pmatrix}
 Q & R \\
S & T
\end{pmatrix}, \label{m0ansatz}
\end{equation}
where $Q,R,S,T$ are $d\times d$ matrices. 
Since $G$ is symmetric,
\begin{gather}
 Q^{T} = Q, \qquad T^{T} = T, \label{ttt}\\
R^{T} = S. \label{rts}
\end{gather}
Now consider equation \eqref{m0eqn},
\begin{equation}
 \begin{pmatrix}
  RQ + QS & R^2 + Q T \\
T Q + S^2 & T R + S T
 \end{pmatrix}
= 
\begin{pmatrix}
0 & 1 \\
1 & 0
\end{pmatrix}
\label{m0matrixeqn}.
\end{equation}
Since we want the duality group element to contain the spacetime metric $g_{ab},$ we assume that $G$ is symmetric and non-degenerate on a $d$-dimensional vector subspace. By an appropriate choice of basis, we can let $T$ be the symmetric, non-degenerate part of $G.$ Now, using the non-degeneracy of $T,$ the matrix equation \eqref{m0matrixeqn} can be solved for $Q$ and $S$:
\begin{gather}
Q = (1-R^2)T^{-1}, \\
S = - T R T^{-1}, \label{Ssoln}
\end{gather}
where $T^{-1}$ is the inverse of $T.$ Using equation \eqref{rts}, the second equation above, \eqref{Ssoln}, can be rewritten
\begin{align}
  R^{T} T^{-1} = - R T^{-1} \quad \implies \quad (R T^{-1})^{T} = - R T^{-1}.
\end{align}
Hence, we define the antisymmetric matrix
\begin{equation}
  B = R T^{-1}.
\end{equation}
Moreover, letting 
\begin{equation}
 T= g^{-1},
\end{equation}
the inverse spacetime metric, we recover the duality group element in the literature, equation \eqref{oddgenmet}. Note that different choices for $R$ and $T$ in terms of $B$ and $g$ correspond to field redefinitions.  

Given the expression for $G$ we can now solve equation \eqref{m1eqn} to find 
 \begin{equation}
 H = 
\begin{pmatrix}
W & X \\
Y & Z
\end{pmatrix},
\end{equation}
where $W, X, Y, Z$ are $d\times d$ matrices, which from equation \eqref{m1eqn}, satisfy
\begin{align}
R W + Q Y + X Q + W S & = 0, \label{m1eqn1} \\
R X + Q Z + X R + W T & = 0, \label{m1eqn2} \\
T W + S Y + Z Q + Y S & = 0, \label{m1eqn3} \\
T X + S Z + Z R + Y T & = 0, \label{m1eqn4}.
\end{align}
The above equations are solved by 
\begin{align}
 W &= - (R X + Q Z + X R) T^{-1} , \label{m1eqn2sol} \\
Y &= - (T X + S Z + Z R) T^{-1}. \label{m1eqn4sol}
\end{align}
Recall, however, that $H$ is symmetric. Hence, $W$ and $Z$ are symmetric, while $$X = Y^{T}.$$
From the above equation  and equation \eqref{m1eqn4sol} we can show that
\begin{equation}
 \tilde{X} = (X - R T^{-1} Z) T^{-1}
\end{equation}
is antisymmetric. Therefore, $H$ is fully determined in terms of $g, B,$ a symmetric matrix $Z$ and an antisymmetric $\tilde{X}$:
\begin{align}
W &= - (B g^{-1} \tilde{X} + \tilde{X} g^{-1} B + g Z g + B Z B), \\
X &=  \tilde{X} g^{-1} + B Z, \\
Y &= - ( g^{-1} \tilde{X} + Z B).
\end{align}

The matrices $Z$ and $\tilde X$ are general symmetric and antisymmetric matrices in terms of the fields of the theory $g$ and $B.$ Note that the dilaton and the determinant of $g$ are described by a scalar density in the duality invariant manifest formulation of the reduced theory. Hence, we expect $H$ to also be independent of the dilaton and the determinant of $g$. Since $\alpha'$ has length dimension two, we also expect  $Z$ and $\tilde{X}$ to have two derivatives. Thus the most general forms of these matrices are\footnote{We exclude terms where the two-form field $B$ is not differentiated, since such terms would lead to manifestly non-gauge invariant terms in the unreduced action.  In any case, such terms would be related to those considered here by non-gauge invariant field redefinitions.}
\begin{gather}
Z^{ab} = a_1 \tilde{g}^{i j} g^{b f} \partial_{i} g^{a e} \partial_{j} g_{e f} + a_2 \tilde{g}^{i j} \left( g^{b f} \partial_{i} g^{a e} \partial_{j} B_{e f} + g^{a f} \partial_{i} g^{b e} \partial_{j} B_{e f} \right) + a_3 \tilde{g}^{i j} g^{a e} g^{b f} g^{g h} \partial_{i} B_{e g} \partial_{j} B_{f h}, \label{Zexp} \\
\tilde{X}_{a b} = a_4 \tilde{g}^{i j} g^{e f} \left(  \partial_{i} g_{a e} \partial_{j} B_{b f} - \partial_{i} g_{b e} \partial_{j} B_{a f} \right).
\label{tildeXexp}
\end{gather}

We leave the coefficients $a_1, a_2, a_3, a_4$ arbitrary, since for any value of these coefficients $\mathcal{G}$ in equation \eqref{Gdef} is a symmetric element of O$(d,d).$ In principle, we are free to choose any value for these constants because different choices correspond to different field redefinitions.  However, as will become clear, a certain choice is preferable in that the higher dimensional action can be written covariantly.

In summary, the duality group element that we consider is
\begin{equation}
\mathcal{G} = 
\begin{pmatrix}
 g - B g^{-1} B & B g^{-1} \\
- g^{-1} B & g^{-1}
\end{pmatrix}
+ \alpha' 
\begin{pmatrix}
- B g^{-1} \tilde{X} - \tilde{X} g^{-1} B - g Z g - B Z B & \tilde{X} g^{-1} + B Z \\
- g^{-1} \tilde{X} - Z B & Z
\end{pmatrix} + \mathcal{O}(\alpha'{}^2), \label{Oddcosel}
\end{equation}
where 
$Z$ and $\tilde{X}$ are given in equations \eqref{Zexp} and \eqref{tildeXexp}.  We stress once more that while the second term is not strictly required, given that it merely corresponds to a field redefinition, it is useful in that it packages the necessary non-gauge invariant field redefinitions that may be required to write the scalar sector of the reduced theory in a duality invariant manner.

\section{Duality completion of Riemann squared term in closed bosonic string theory}
\label{secbos}

In this section we apply the arguments of section \ref{sec:dualcomp} to the Riemann squared $\alpha'$ term in closed bosonic string theory. The $\alpha'$ corrections to the low energy effective theory of closed bosonic string theory at tree-level can be found by string scattering amplitude calculations \cite{grosssloan, metsaevtseytlin}, or equivalently \cite{CFMP, lovelace84, sen85} it is given by the two-loop vanishing of the closed bosonic string sigma-model $\beta$-functionals \cite{metsaevtseytlin}. However, here we assume that the only known $\alpha'$ correction to the effective action is Riemann squared. For various reasons it has been argued in the literature \cite{zwiebach, grosssloan} that the Riemann squared should come as part of a Lovelock term. However, this term is related to Riemann squared by field redefinition. Therefore, consider    
\begin{equation}
 S_{\textup{bosonic}} = \int \d x^{\mu} \sqrt{g} \e^{-2 \phi} \left( R + 4 g^{\mu \nu} (\partial_{\mu} \phi) (\partial_{\nu} \phi) - \frac{1}{12} H^{2} + \alpha' \lambda_{0}  R^{\mu \nu \rho \sigma} R_{\mu \nu \rho \sigma}\right),
\end{equation}
where here $\mu, \nu, \dots = 1, \dots, 26$ and $\lambda_0=1/4;$
\begin{equation}
 H_{\mu \nu \rho} = 3 \partial_{[\mu} B_{\nu \rho]}
\end{equation}
is the field strength of the 2-form of closed bosonic string theory and $\phi$ is the dilaton.

The scalar part of the reduction of $$ S^{(0)}= \int \d x^{\mu} \sqrt{g} \e^{-2 \phi} \left( R + 4 g^{\mu \nu} (\partial_{\mu} \phi) (\partial_{\nu} \phi) - \frac{1}{12} H^{2} \right)$$ is given, up to integration by parts by
\begin{align}
S^{(0)}_{\textup{red}_S} &= \int \d \tilde{x}^{i} \sqrt{\tilde{g}} \sqrt{\hat{g}} \e^{-2 \phi} \tilde{g}^{ij} \left( \frac{1}{4} \partial_{i} \hat{g}^{ab} \partial_{j} \hat{g}_{ab} + \frac{1}{4} \partial_{i} \textup{log}\, \hat{g} \partial_{j} \textup{log} \,\hat{g}\notag \right. \\ 
& \qquad \qquad \qquad \qquad \qquad \qquad \qquad \qquad \quad \left.  - 2 \partial_{i} \phi \partial_{j} \textup{log}\, \hat{g} + 4 \partial_{i} \phi \partial_{j} \phi -\frac{1}{4} \hat{g}^{ab} \hat{g}^{cd} \partial_{i} B_{ac} \partial_{j} B_{bd} \right), \label{2devscal}
\end{align}
where we have used equations \eqref{riemscal1}--\eqref{riemscal7} in appendix \ref{dimred} for the action of the scalars in the reduction of the Einstein-Hilbert term. 
Meanwhile, using the results of appendix \ref{dimred}, the scalar part of the reduction of the $\alpha'$ correction is 
\begin{align}
 R^{\mu \nu \rho \sigma}  R_{\mu \nu \rho \sigma} \Bigl|_{\textup{scalars}}
&= - \tilde{g}^{ij} \tilde{g}^{kl}  \left( \partial_{i} \partial_{k} \hat{g}^{ab} \partial_{j} \partial_{l} \hat{g}_{ab} + \hat{g}^{ab}  \partial_{i} \partial_{k} \hat{g}_{ac} \partial_{j} \hat{g}^{cd} \partial_{l} \hat{g}_{bd} - \frac{1}{8} \partial_{i} \hat{g}^{ab} \partial_{k} \hat{g}_{ab} \partial_{j} \hat{g}^{cd} \partial_{l} \hat{g}_{cd}  \right. \notag\\
& \left. \qquad \qquad \qquad \qquad  - \frac{1}{2} \partial_{i} \hat{g}^{ab} \partial_{j} \hat{g}_{bc} \partial_{k} \hat{g}^{cd} \partial_{l} \hat{g}_{da}  + \frac{3}{8} \partial_{i} \hat{g}^{ab} \partial_{k} \hat{g}_{bc} \partial_{j} \hat{g}^{cd} \partial_{l} \hat{g}_{da} \right). \label{riem2redscal}
\end{align}
For ease of reading, we will simply denote the internal metric $\hat{g}$ by $g$ in the rest of this section.   

Since duality in the reduced theory dictates that the scalars parametrise the duality coset element, we must be able to write the terms in \eqref{riem2redscal} in terms of the O$(d,d)$/O$(d)\times$O$(d)$ coset element derived in section \ref{secOddcoset}, expression \eqref{Oddcosel}, and the invariant measure defined in equation \eqref{measinv}. As we are only working to first order in $\alpha',$ for the terms in equation \eqref{riem2redscal} the only relevant part of the coset element in equation \eqref{Gdef} is 
\begin{equation}
 G_{AB} = 
\begin{pmatrix}
 g_{ab} + g^{ef} B_{a e}  B_{bf} & B_{a e} g^{e d} \\
- g^{c e} B_{e b} & g^{c d}
\end{pmatrix}. \label{m02}
\end{equation}
The inverse of $G,$ from equation \eqref{m0eqn}, is
\begin{align}
 G^{AB} &= \eta^{A C} G_{CD} \eta^{DB} \label{Ginveqn} \\
&= 
\begin{pmatrix}
 g^{ab} &  - g^{a e} B_{e d} \\
B_{c e} g^{e b} &  g_{cd} + g^{ef} B_{c e}  B_{df}
\end{pmatrix}. \label{invm0}
\end{align}
Recall that the determinant of the internal metric and the dilaton appear in the invariant combination 
\begin{equation}
 \e^{-2d} = \sqrt{g} \e^{-2 \phi}.
\label{ddef}
\end{equation}
 
It is straightforward to see that 
\begin{align}
 \tilde{g}^{ij} \tilde{g}^{kl} \partial_{i} \partial_{k} g^{ab} \partial_{j} \partial_{l} g_{ab} &=  \frac{1}{2} \tilde{g}^{ij} \tilde{g}^{kl} \partial_{i} \partial_{k} G^{AB} \partial_{j} \partial_{l} G_{AB} \Big|_{B=0}, \\
 \tilde{g}^{ij} \tilde{g}^{kl} \partial_{i} g^{ab} \partial_{k} g_{ab} \partial_{j} g^{cd} \partial_{l} g_{cd} &= \frac{1}{4}  \tilde{g}^{ij} \tilde{g}^{kl} \partial_{i} G^{AB} \partial_{k} G_{AB} \partial_{j} G^{CD} \partial_{l} G_{CD} \Big|_{B=0}, \\
\tilde{g}^{ij} \tilde{g}^{kl} \partial_{i} g^{ab} \partial_{j} g_{bc} \partial_{k} g^{cd} \partial_{l} g_{da} &= \frac{1}{2} \tilde{g}^{ij} \tilde{g}^{kl} \partial_{i} G^{AB} \partial_{j} G_{BC} \partial_{k} G^{CD} \partial_{l} G_{DA}\Big|_{B=0}, \\ 
\tilde{g}^{ij} \tilde{g}^{kl} \partial_{i} g^{ab} \partial_{k} g_{bc} \partial_{j} g^{cd} \partial_{l} g_{da} &= \frac{1}{2} \tilde{g}^{ij} \tilde{g}^{kl} \partial_{i} G^{AB} \partial_{k} G_{BC} \partial_{j} G^{CD} \partial_{l} G_{DA} \Big|_{B=0}.
\end{align}
However, note that the second term in equation \eqref{riem2redscal} is not expressible in terms of $G.$ For example,
\begin{align}
\tilde{g}^{ij} \tilde{g}^{kl} G^{AB}  \partial_{i} \partial_{k} G_{AC} \partial_{j} G^{CD} \partial_{l} G_{BD}\Big|_{B=0} &= \tilde{g}^{ij} \tilde{g}^{kl} \left( g^{ab}  \partial_{i} \partial_{k} g_{ac} \partial_{j} g^{cd} \partial_{l} g_{bd} + g_{ab}  \partial_{i} \partial_{k} g^{ac} \partial_{j} g_{cd} \partial_{l} g^{bd} \right), \notag \\
&= \tilde{g}^{ij} \tilde{g}^{kl} \partial_{i} g^{ab} \partial_{k} g_{bc}\left( \partial_{j} g^{cd} \partial_{l} g_{da} -  \partial_{l} g^{cd} \partial_{j} g_{da} \right).
\end{align}
The reason for the fact that the second term in equation \eqref{riem2redscal} cannot be expressed in terms of $G$ is essentially equation \eqref{Ginveqn}. If there are an odd number of coset elements that are differentiated, there is a relative minus sign between raising (or lowering) the coset elements that are differentiated with $G^{-1}$ (or $G$) and with $\eta.$ This observation was made in the reduction of the closed bosonic string effective action with $\alpha'$ corrections to one dimension by Meissner \cite{meissner} and more recently in the context of double field theory \cite{hohmzwiebach1}. In the reduction to one dimension, Meissner showed that these terms can be obtained from the rewriting of the two-derivative terms in the scalar part of the reduced theory action in terms of a modified coset element. 

In section \ref{sec:dualcomp}, it was shown that the expression for the two-derivative, scalar part of the reduced action, \eqref{2devscal}, can be written as
\begin{equation}
S^{(0)}_{\textup{red}_S}= \int \d \tilde{x}^{i} \sqrt{\tilde{g}} \tilde{g}^{ij} \e^{-2 d}  \left( \frac{1}{8} \partial_{i} G^{AB} \partial_{j} G_{AB} + 4 \partial_{i} d \partial_{j} d \right). \label{2devscalG}
\end{equation} 
Now, we replace the zeroth-order coset element $G$ in the above expression with the full coset element $\mathcal{G}.$  Using equation \eqref{Gdef}, 
\begin{equation}
 \frac{1}{8} \tilde{g}^{ij} \partial_{i} \mathcal{G}^{AB} \partial_{j} \mathcal{G}_{AB} = \frac{1}{8} \tilde{g}^{ij} \partial_{i} G^{AB} \partial_{j} G_{AB} +  \frac{1}{4} \alpha' \tilde{g}^{ij} \partial_{i} H^{AB} \partial_{j} G_{AB} + \mathcal{O}(\alpha'^2). 
\label{2devfullcoset}
\end{equation}
In the above expression 
\begin{equation}
 \mathcal{G}^{AB} = G^{AB} + \alpha' H^{AB}  + \mathcal{O}(\alpha'^2)
\end{equation}
is the inverse of $\mathcal{G}_{AB}.$ However, note that 
\begin{equation}
 H^{AB} = \eta^{AC} H_{CD} \eta^{DB} 
\end{equation}
is \emph{not} the inverse of $H_{AB}.$

Consider the term that is first-order in $\alpha'$ on the right-hand-side of equation \eqref{2devfullcoset}. Using equation \eqref{Oddcosel},
\begin{align}
\tilde{g}^{ij} \partial_{i} H^{AB} \partial_{j} G_{AB}
&= \tilde{g}^{ij} \left( 2 \partial_{i} Z^{ab} \partial_{j} g_{ab}  - 2 Z^{ab} g_{ac} \partial_{i} g_{bd} \partial_{j} g^{cd}  \notag \right.\\
&\hspace*{35mm} \left.  
- 2 g^{ab} g^{cd} \partial_{i} \tilde{X}_{ac} \partial_{j} B_{bd} - 2 g^{ab} Z^{cd} \partial_{i} B_{ac} \partial_{j} B_{bd} \right).
\label{1order}
\end{align}
Substituting equation \eqref{Zexp} in the expression above we obtain
\begin{align}
 \tilde{g}^{ij} \partial_{i} H^{AB} \partial_{j} G_{AB} \Big|_{B=0} = 4 a_1 \tilde{g}^{ij}  \tilde{g}^{k l} \left( g^{ab}  \partial_{i} \partial_{k} g_{ac} \partial_{j} g^{cd} \partial_{l}  g_{bd} + \partial_{i} g^{ab} \partial_{(j|} g_{bc} \partial_{|k)} g^{cd} \partial_{l} g_{da}  \right),
\label{2devfullG1}
\end{align}
where we have ignored a derivative on $\tilde{g}^{ij}$ in the above because this term does not contribute to the scalar sector of the reduced theory. Comparing equation \eqref{2devfullcoset} and the equation above with equation \eqref{riem2redscal}, we see that
\begin{equation}
 a_1 = - \lambda_0.
 \label{a1val}
\end{equation}
Therefore, in summary
\begin{align}
 \left(R + \alpha' \lambda_0 R^{\mu \nu \rho \sigma}  R_{\mu \nu \rho \sigma} \right) \Bigl|_{\textup{scalars}} = L_{\textup{red}_S}\Big|_{B, \phi=0},
\label{gravredscal}
\end{align}
where
\begin{align}
L_{\textup{red}_S} =
  \tilde{g}^{ij} & \left( \frac{1}{8}  \partial_{i} \mathcal{G}^{AB} \partial_{j} \mathcal{G}_{AB} + 4  \partial_{i} d \partial_{j} d \right) \notag \\
   & - \frac{1}{2} \alpha' \lambda_0 \tilde{g}^{ij} \tilde{g}^{kl} \left( \partial_{i} \partial_{k} \mathcal{G}^{AB} \partial_{j} \partial_{l} \mathcal{G}_{AB} 
 - \frac{1}{16} \partial_{i} \mathcal{G}^{AB} \partial_{k} \mathcal{G}_{AB} \partial_{j} \mathcal{G}^{CD} \partial_{l} \mathcal{G}_{CD} \right. \notag \\
 & \hspace{30mm} \left. - \partial_{i} \mathcal{G}^{AB} \partial_{j} \mathcal{G}_{BC} \partial_{k} \mathcal{G}^{CD} \partial_{l} \mathcal{G}_{DA}  - \frac{1}{8}  \partial_{i} \mathcal{G}^{AB} \partial_{k} \mathcal{G}_{BC} \partial_{j} \mathcal{G}^{CD} \partial_{l} \mathcal{G}_{DA} \right).
 \label{Lredscal}
\end{align}

Now that we have written the gravitational part of the scalar sector of the reduced theory in terms of the O$(d,d)$ coset element we can evaluate the right hand side of the above equation, \eqref{Lredscal}, to find the dependence of the scalar part of the reduced lagrangian on the 2-form $B$ and the dilaton $\phi.$

We have already established that the zeroth order term in $\alpha'$ in the first two expressions on the right hand side of equation \eqref{Lredscal} reduce to the lagrangian in equation \eqref{2devscalG} and thus reproduces the scalar sector of the two-derivative part of the theory. We now evaluate the first order term, which using equation \eqref{2devfullcoset} is 
\begin{align}
  \frac{1}{4} \tilde{g}^{ij} \partial_{i} H^{AB} \partial_{j} G_{AB} = & - \lambda_0 \tilde{g}^{ij}  \tilde{g}^{k l} \left( g^{ab}  \partial_{i} \partial_{k} g_{ac} \partial_{j} g^{cd} \partial_{l}  g_{bd} + \partial_{i} g^{ab} \partial_{(j|} g_{bc} \partial_{|k)} g^{cd} \partial_{l} g_{da} \right) \notag \\
  & + a_2 \tilde{g}^{ij}  \tilde{g}^{k l} \left(  \partial_{i} g_{a b} \partial_{j} g^{a d} \partial_{k} g^{b c} \partial_{l} B_{c d} + g^{a b} \partial_{i} \partial_{k} g_{a c} \partial_{j} g^{c d} \partial_{l} B_{b d} \right. \notag \\
  &\hspace{65mm} \left. - g^{a b} \partial_{i} \partial_{k} B_{a c} \partial_{j} g_{b d} \partial_{l} g^{c d} \right) \notag \\
  & + \frac{1}{2} \tilde{g}^{ij}  \tilde{g}^{k l} \left( (a_3 + \lambda_0) g^{a b} g^{c d} \partial_{i} g^{ef} \partial_{j} g_{a e} \partial_{k} B_{f c} \partial_{l} B_{b d} - a_3 \partial_{i} g^{a b} \partial_{j} g^{c d} \partial_{k} B_{a c} \partial_{l} B_{b d}\right. \notag \\
  & \hspace{25mm} - 2 a_4 g^{a b} g^{c d} \partial_{i} g^{e f} \partial_{k} g_{ae} \partial_{j} B_{bd} \partial_{l} B_{cf} \notag \\
  & \hspace{35mm} - 2 a_4 g^{a b} g^{c d} g^{e f} \partial_{i} \partial_{k} g_{ae} \partial_{j} B_{bd} \partial_{l} B_{cf} \notag \\
  & \hspace{45mm} \left. - 2 (a_3 + a_4) g^{a b}  \partial_{i} \partial_{k} B_{a c} \partial_{j} g^{c d} \partial_{l} B_{b d} \notag \right) \\
  & - a_2 \tilde{g}^{ij}  \tilde{g}^{k l} g^{a b} g^{c d} \partial_{i} g^{e f} \partial_{j} B_{e d} \partial_{k} B_{a f} \partial_{l} B_{b c} \notag \\
  & - \frac{1}{2} a_3 \tilde{g}^{ij}  \tilde{g}^{k l} g^{a b} g^{c d} g^{e f} g^{g h} \partial_{i} B_{ac} \partial_{j} B_{be} \partial_{k} B_{dg} \partial_{l} B_{fh}, 
  \label{HGfull}
\end{align}
where we have used equations \eqref{1order}, \eqref{Zexp}, \eqref{tildeXexp} and \eqref{a1val}.

Since the rest of the terms in equation \eqref{Lredscal} are already at order $\alpha'$, the order to which we are working to, we can simply replace $\mathcal{G}$ by $G$ in these terms
\begin{align}
  - \frac{1}{2} \alpha' \lambda_0 \tilde{g}^{ij} \tilde{g}^{kl} & \left( \partial_{i} \partial_{k} G^{AB}  \partial_{j} \partial_{l} G_{AB}  - \frac{1}{16} \partial_{i} G^{AB} \partial_{k} G_{AB} \partial_{j} G^{CD} \partial_{l} G_{CD} \right. \notag \\
 & \hspace{15mm} \left. - \partial_{i} G^{AB} \partial_{j} G_{BC} \partial_{k} G^{CD} \partial_{l} G_{DA}  - \frac{1}{8}  \partial_{i} G^{AB} \partial_{k} G_{BC} \partial_{j} G^{CD} \partial_{l} G_{DA} \right).
 \label{othertermsG}
\end{align}
Furthermore, to evaluate these terms we first prove that these terms have no bare $B$ terms. 

It is straightforward to verify that 
\begin{align}
G_{AB} = L_{A}{}^{C} L_{B}{}^{D} D_{CD}, \label{Gvielbein} \\
G^{AB} = E_{C}{}^{A} E_{D}{}^{B} D^{CD},
\label{Ginvvielbein}
\end{align}
where 
\begin{align}
 L_{A}{}^{B} &= 
\begin{pmatrix}
 \delta_{a}{}^{b} & B_{a d} \\
0 & \delta^{c}{}_{d} 
\end{pmatrix}, \qquad \qquad 
 E_{A}{}^{B} = 
\begin{pmatrix}
 \delta_{a}{}^{b} & - B_{a d} \\
0 & \delta^{c}{}_{d} 
\end{pmatrix}, \label{LexpEexp} \\
D_{AB} &= 
\begin{pmatrix}
 g_{a b} & 0 \\
0 & g^{c d} 
\end{pmatrix}, \qquad \qquad \
 D^{AB} = 
\begin{pmatrix}
 g^{ab} & 0 \\
0 & g_{cd} 
\end{pmatrix}. \label{Dexp}
\end{align}
Or, if we denote the indices $A= (a, \tilde{a})$ and $B= (b, \tilde{b}),$
\begin{gather} 
L_{a}{}^{b} = E_{a}{}^{b} = \delta_{a}^{b} , \label{LEtopdiag} \\ 
L_{\tilde{a}}{}^{\tilde{b}} = E_{\tilde{a}}{}^{\tilde{b}} = \delta_{\tilde{b}}^{\tilde{a}}, \\ 
L_{a}{}^{\tilde{b}} = - E_{a}{}^{\tilde{b}} = B_{a \tilde{b}}, \label{LEB}
\\ L_{\tilde{a}}{}^{b} = E_{\tilde{a}}{}^{b} =  0. \label{LEzero}
\end{gather}
Note that 
\begin{equation}
 L_{A}{}^{C} E_{C}{}^{B} = \delta_{A}^{B}, \qquad  L_{C}{}^{A} E_{B}{}^{C} = \delta_{B}^{A}.
\label{LEeqn}
\end{equation}

Consider the terms in expression \eqref{othertermsG}. Note that for these terms the group element $G$ indices only contract with the indices of the inverse group element. From equations \eqref{Gvielbein} and \eqref{Ginvvielbein} this implies that when the group element and its inverse are written in terms of $L$ and $E,$ the lower index of the $L$ always contracts with an upper index on $E.$ Let us now assume for contradiction that there is a bare $B,$ i.e.\ the two-form $B$ is not differentiated. From equations \eqref{LexpEexp} and \eqref{Dexp}, it is clear that in such a case there must be either a bare $L$ or $E.$ First consider the former case. Recall that the lower index of the $L$ contracts with the upper index of an $E$
\begin{equation}
 L_{A}{}^{B} \partial E_{C}{}^{A} \quad \textup{or} \quad  L_{A}{}^{B} \partial \partial E_{C}{}^{A}.
\end{equation}
The $E$ must be differentiated otherwise from equation \eqref{LEeqn} the $L$ and $E$ would contract to a Kronecker delta symbol, contradicting the assumption that there is a bare $B.$ From equation \eqref{LEB}, there must, in particular, be a term of the form
\begin{equation}
 L_{a}{}^{\tilde{b}} \partial E_{C}{}^{a} \quad \textup{or} \quad  L_{a}{}^{\tilde{b}} \partial \partial E_{C}{}^{a}.
\end{equation}
From equations \eqref{LEtopdiag} and \eqref{LEzero}, $E$ is either a Kronecker delta symbol or zero. In either case the expression above vanishes. Similarly, if a bare $B$ term arises from a bare $E,$ the term must contain the expression 
\begin{equation}
 E_{a}{}^{\tilde{b}} \partial L_{C}{}^{a} \quad \textup{or} \quad E_{a}{}^{\tilde{b}} \partial \partial L_{C}{}^{a}.
\end{equation}
By the argument above, this expression also vanishes contradicting the initial assumption that there is a term with a bare $B.$ Hence we have shown that all the two-form fields are differentiated in expression \eqref{othertermsG}. So we can simplify the calculation of the terms  by writing  
\begin{gather}
 \partial_{i} \partial_{j} G_{AB} = 
\begin{pmatrix}
 \partial_{i} \partial_{j} g_{ab} + 2 g^{ef} \partial_{(i|} B_{a e}  \partial_{|j)} B_{bf} & g^{e d} \partial_{i} \partial_{j} B_{a e} + 2 \partial_{(i|} g^{e d} \partial_{|j)} B_{a e} \\
- g^{c e} \partial_{i} \partial_{j} B_{e b} - 2 \partial_{(i|} g^{c e}  \partial_{|j)} B_{e b} & \partial_{i} \partial_{j} g^{c d}
\end{pmatrix}, \label{G2deriv} \\
 \partial_{i} \partial_{j} G^{AB} = 
\begin{pmatrix}
\partial_{i} \partial_{j} g^{a b}  & - g^{a e} \partial_{i} \partial_{j} B_{e d} - 2 \partial_{(i|} g^{a e} \partial_{|j)} B_{e d} \\
 g^{e b} \partial_{i} \partial_{j} B_{c e} + 2 \partial_{(i|} g^{e b}  \partial_{|j)} B_{c e} & \partial_{i} \partial_{j} g_{c d} + 2 g^{ef} \partial_{(i|} B_{c e}  \partial_{|j)} B_{d f}
\end{pmatrix}, \label{Ginv2deriv} \\
 \partial_{i} G_{AB} = 
\begin{pmatrix}
 \partial_{i} g_{ab} & g^{e d} \partial_{i} B_{a e}  \\
- g^{c e} \partial_{i} B_{e b} & \partial_{i} g^{c d}
\end{pmatrix}, \hspace{10mm}
 \partial_{i} G^{AB} = 
\begin{pmatrix}
\partial_{i} g^{ab} &  - g^{a e} \partial_{i}B_{e d} \\
g^{e b} \partial_{i} B_{c e}  &  \partial_{i} g_{cd} 
\end{pmatrix} \label{Ginvderiv}
\end{gather}
in expression \eqref{othertermsG}.
It is now straightforward to show that 
\begin{align}
 \tilde{g}^{ij} \tilde{g}^{kl} \partial_{i} \partial_{k} G^{AB} \partial_{j} \partial_{l} G_{AB} &= 2 \tilde{g}^{ij} \tilde{g}^{kl} \left( \partial_{i} \partial_{k} g^{ab} \partial_{j} \partial_{l} g_{ab} + 2 g^{a b} \partial_{i} \partial_{k} g^{c d} \partial_{j} B_{a c} \partial_{l} B_{b d} \right. \notag \\
 & \hspace{20mm} - 4 \partial_{i} g^{a b} \partial_{(j} g^{c d} \partial_{k)} B_{a c} \partial_{l} B_{b d} - 4 g^{a b} \partial_{i} g^{c d} \partial_{k} B_{a c} \partial_{j} \partial_{l} B_{b d} \notag \\
  & \hspace{40mm} \left. - g^{a b} g^{c d} \partial_{i} \partial_{k} B_{a c} \partial_{j} \partial_{l} B_{b d} \right), 
  \label{ddgddgfull}
\end{align}
\begin{align}
  \tilde{g}^{ij} \tilde{g}^{kl} \partial_{i} G^{AB} \partial_{k} G_{AB} \partial_{j} G^{CD} \partial_{l} G_{CD} &= 4 \tilde{g}^{ij} \tilde{g}^{kl} \left(  \partial_{i} g^{ab} \partial_{k} g_{ab} \partial_{j} g^{cd} \partial_{l} g_{cd} - 2 g^{cd} g^{ef} \partial_{i} g^{ab} \partial_{k} g_{ab} \partial_{j} B_{ce} \partial_{l} B_{df} \right. \notag \\
   & \hspace{30mm} \left. + g^{a b} g^{cd} g^{ef} g^{g h} \partial_{i} B_{a c} \partial_{j} B_{e g} \partial_{k} B_{b d} \partial_{l} B_{f h} \right),
\end{align}
\begin{align}
  \tilde{g}^{ij} \tilde{g}^{kl} \partial_{i} G^{AB} \partial_{j} G_{BC} \partial_{k} G^{CD} \partial_{l} G_{DA} &= 2 \tilde{g}^{ij} \tilde{g}^{kl} \left( \partial_{i} g^{ab} \partial_{j} g_{bc} \partial_{k} g^{cd} \partial_{l} g_{da} + 2 \partial_{i} g^{a b} \partial_{j} B_{b c} \partial_{k} g^{c d} \partial_{l} B_{d a} \right. \notag \\
  & \hspace{20mm}- 4 g^{a b} g^{c d} \partial_{i} g^{e f} \partial_{(j|} g_{f a} \partial_{|k)} B_{e c} \partial_{l}  B_{b d} \notag \\
  & \hspace{30mm} \left. + g^{a b} g^{cd} g^{ef} g^{g h} \partial_{i} B_{a c} \partial_{j} B_{d e} \partial_{k} B_{f g} \partial_{l} B_{h b} \right),
\end{align}
\begin{align}
  \tilde{g}^{ij} \tilde{g}^{kl} \partial_{i} G^{AB} \partial_{k} G_{BC} \partial_{j} G^{CD} \partial_{l} G_{DA} &= 2 \tilde{g}^{ij} \tilde{g}^{kl} \left( \partial_{i} g^{ab} \partial_{k} g_{bc} \partial_{j} g^{cd} \partial_{l} g_{da} + 2 \partial_{i} g^{a b} \partial_{k} B_{b c} \partial_{j} g^{c d} \partial_{l} B_{d a} \right. \notag \\
  & \hspace{20mm}- 4 g^{a b} g^{c d} \partial_{i} g^{e f} \partial_{k} g_{f a} \partial_{j} B_{b c} \partial_{l}  B_{e d} \notag \\
  & \hspace{30mm} \left. + g^{a b} g^{cd} g^{ef} g^{g h} \partial_{i} B_{a c} \partial_{k} B_{d e} \partial_{j} B_{f g} \partial_{l} B_{h b} \right).
  \label{d4g3}
\end{align}

Let us now consider the order $\alpha'$ terms in equation \eqref{Lredscal}, which are given by equations \eqref{HGfull} and \eqref{ddgddgfull}--\eqref{d4g3}. These terms were found by requiring that at zeroth order in the field $B$ they produce the scalar sector of the reduced Riemann squared term. The only contributions at odd order in the field $B$ come from terms in  
$$\tilde{g}^{ij} \partial_{i} H^{AB} \partial_{j} G_{AB}$$
with coefficient $a_2,$ see equation \eqref{HGfull}. These terms cannot be made covariant, hence we set $$a_2=0.$$ 
Note that for any value of $a_2$ the scalars parametrise the duality coset element and so any value is a valid choice. However, only for some value will the duality coset element give rise to a covariant description. Other choices will be related to this by non-gauge invariant field redefinitions.

Now, consider terms quartic in the field $B,$
\begin{align} \label{alphapB4}
 L_{\textup{red}_S}\Big|_{\mathcal{O}(\alpha', B^4)} = \tilde{g}^{ij} \tilde{g}^{k l} g^{a b} g^{c d} g^{ef} g^{gh}
  &\left\{ \left(\lambda_0- \textstyle{\frac{a_3}{2}}\right) \partial_{i} B_{ae} \partial_{j} B_{bg} \partial_{k} B_{cf} \partial_{l} B_{dh} 
  + \textstyle{\frac{1}{8}} \lambda_0 \partial_{i} B_{ac} \partial_{j} B_{eg} \partial_{k} B_{bd} \partial_{l} B_{fh} \right. \notag \\[1mm]
   & \hspace{5mm} \left. +\textstyle{\frac{1}{8}} \lambda_0 \partial_{i} B_{ac} \partial_{j} B_{eg} \partial_{k} B_{bf} \partial_{l} B_{dh} \right\}.
\end{align}
Clearly, a gauge invariant term that gives rise to such terms under reduction must be, schematically, of the form
\begin{equation*}
 H \, H \, H \, H
\end{equation*}
with various contractions.  Of course, the precise nature of the contractions can be inferred from the nature of the contractions in the terms above.  In particular, it is clear to see that the terms in the first line should come from a term of the form
\begin{equation}
 H^{\mu \nu \tau} H_{\mu \nu \eta} H^{\rho \sigma \eta} H_{\rho \sigma \tau} \longrightarrow
 \tilde{g}^{ij} \tilde{g}^{k l} g^{a b} g^{c d} g^{ef} g^{gh} \left(4 \partial_{i} B_{ae} \partial_{j} B_{bg} \partial_{k} B_{cf} \partial_{l} B_{dh} + \partial_{i} B_{ac} \partial_{j} B_{eg} \partial_{k} B_{bd} \partial_{l} B_{fh}
 \right),
\end{equation}
where indices on the left hand side have been raised using the inverse metric $g^{\mu \nu}$ and the arrow indicates the scalars that are obtained under a reduction of the term.  Hence, we conclude that
\begin{equation}
 \frac{\lambda_0}{8}  H^{\mu \nu \tau} H_{\mu \nu \eta} H^{\rho \sigma \eta} H_{\rho \sigma \tau}
\end{equation}
and only this term can account for the first two terms on the right hand side of equation \eqref{alphapB4}, provided that we choose
\begin{equation} \label{a3}
 a_3=\lambda_0.
\end{equation}
As emphasised earlier, we are free to make such a choice and any other choice will be related to this via a non-gauge invariant transformation.

Similarly, the final term in equation \eqref{alphapB4} can only come from a term of the form
\begin{equation}
 {H_{\mu}}^{\sigma \tau} {H_{\rho}}^{\mu \nu} {H^{\rho}}_{\tau \eta} {H^{\eta}}_{\nu \sigma} \longrightarrow
 3 \tilde{g}^{ij} \tilde{g}^{k l} g^{a b} g^{c d} g^{ef} g^{gh} \partial_{i} B_{ac} \partial_{j} B_{eg} \partial_{k} B_{bf} \partial_{l} B_{dh}.
\end{equation}
Hence, the final term in equation \eqref{alphapB4} comes from a reduction of the following and only the following term
\begin{equation}
 \frac{\lambda_0}{24}  {H_{\mu}}^{\sigma \tau} {H_{\rho}}^{\mu \nu} {H^{\rho}}_{\tau \eta} {H^{\eta}}_{\nu \sigma}.
\end{equation}
In summary,
\begin{equation}
 L_{\textup{red}_S}\Big|_{\mathcal{O}(\alpha', B^4)} =\lambda_0 \left(\frac{1}{8}  H^{\mu \nu \tau} H_{\mu \nu \eta} H^{\rho \sigma \eta} H_{\rho \sigma \tau} + \frac{1}{24}  {H_{\mu}}^{\sigma \tau} {H_{\rho}}^{\mu \nu} {H^{\rho}}_{\tau \eta} {H^{\eta}}_{\nu \sigma} \right)\Big|_{\textup{scalars}}.
\end{equation}

The attentive reader may point out that the form of terms quartic in the field $B$ is constrained to the extent that there are only four such possible terms: the three terms appearing in the equations above in addition to a term of the form
\begin{equation*}
(\tilde{g}^{ij} g^{a b} g^{c d} \partial_{i} B_{ac} \partial_{j} B_{bd}) (\tilde{g}^{k l} g^{ef} g^{gh} \partial_{k} B_{eg} \partial_{l} B_{fh}).
\end{equation*}
The term above is very simple to take care of since it will come from a reduction of a term of the form
\begin{equation*}
 (H^{\mu \nu \rho} H_{\mu \nu \rho})(H^{\tau \sigma \eta} H_{\tau \sigma \eta}). 
\end{equation*}
This as well as the fact that in deriving the result above we essentially had three equations for three unknowns may tempt the reader to conclude that the fact that we were able to assemble terms of order four in $B$ into gauge invariant terms in the full theory is unsurprising.  However, the result is not as trivial as it may seem. First, the result proved above that there are no bare $B$'s is crucial in deriving this result.  Otherwise the task of finding gauge invariant terms in the full theory that reduce to such terms would not be such a nice one.  In addition, the form of the eventual equations need not have been so.  The fact that the $\alpha'$ correction to the duality group element contributes in the way that it does is crucial from a practical point of view. Of course, there is also the fact that the particular form of the gauge invariant terms in the full theory ought to be consistent with previous results in the literature.

Finally, we are left with second order terms in the field $B.$  Substituting the value of $a_3,$ equation \eqref{a3}, these are of the form
\begin{align}
L_{\textup{red}_S}\Big|_{\mathcal{O}(\alpha', B^2)} = 
  \tilde{g}^{ij} \tilde{g}^{k l} &  \left\{ 
  \lambda_0 g^{a b} g^{c d} \partial_{i} \partial_{k} B_{a c} \partial_{j} \partial_{l} B_{b d} + (3 \lambda_0 -a_4) g^{a b}  \partial_{i} \partial_{k} B_{a c} \partial_{j} g^{c d} \partial_{l} B_{b d} \right. \notag \\
  &\ \ + (a_4 + 2 \lambda_0) g^{a b} g^{c d} g^{e f} \partial_{i} \partial_{k} g_{a c} \partial_{j} B_{b e} \partial_{l} B_{d f} \notag \\[2pt] 
  &\ \ - \lambda_0  g^{a b} g^{c d} \partial_{i} g^{ef} \partial_{j} g_{a e} \partial_{k} B_{f c} \partial_{l} B_{b d}
  - \frac{\lambda_0}{4}  g^{cd} g^{ef} \partial_{i} g^{ab} \partial_{k} g_{ab} \partial_{j} B_{ce} \partial_{l} B_{df} \notag \\[3pt]
  &\ \ \left.+ \frac{5\lambda_0}{4} \partial_{i} g^{a b} \partial_{j} g^{c d} \partial_{k} B_{a c} \partial_{l} B_{b d}  +\left(a_4 +\frac{3\lambda_0}{2} \right) g^{a b} g^{c d} \partial_{i} g^{e f} \partial_{k} g_{ae} \partial_{j} B_{bc} \partial_{l} B_{fd} \right\}.
\end{align}
The terms above in the first line of the expression on the right hand side can only come from the reduction of a term of the form\footnote{This is strictly not true. However, any other term giving rise to such a term under reduction will be related to this term by use of the Bianchi identity $\nabla_{[\mu}H_{\nu \rho \sigma]}=0.$  Thus, in this sense the term is unique.}
\begin{align}
 \frac{1}{3} \nabla^\mu H^{\nu \rho \sigma} \nabla_\mu H_{\nu \rho \sigma} \longrightarrow 
 \tilde{g}^{ij} \tilde{g}^{k l} &\left\{ g^{a b} g^{c d} \partial_{i} \partial_{k} B_{a c} \partial_{j} \partial_{l} B_{b d} + 2 g^{a b}  \partial_{i} \partial_{k} B_{a c} \partial_{j} g^{c d} \partial_{l} B_{b d} \right. \notag \\
 &\ \ - g^{a b} g^{c d} \partial_{i} g^{ef} \partial_{j} g_{a e} \partial_{k} B_{f c} \partial_{l} B_{b d} 
 + \frac{1}{2} \partial_{i} g^{a b} \partial_{j} g^{c d} \partial_{k} B_{a c} \partial_{l} B_{b d} \notag \\
 &\ \ \left. + g^{a b} g^{c d} \partial_{i} g^{e f} \partial_{k} g_{ae} \partial_{j} B_{bc} \partial_{l} B_{fd}
 - \frac{1}{4}  g^{cd} g^{ef} \partial_{i} g^{ab} \partial_{k} g_{ab} \partial_{j} B_{ce} \partial_{l} B_{df} \right\},
\end{align}
where $\nabla$ denotes the covariant derivative induced by metric $g_{\mu \nu}$. Comparing the coefficient of the second term, we deduce that this is only possible for
\begin{equation}
 a_4=\lambda_0.
\end{equation}
Hence,
\begin{align}
L_{\textup{red}_S}\Big|_{\mathcal{O}(\alpha', B^2)} = & \lambda_0 \left( \frac{1}{3} \nabla^\mu H^{\nu \rho \sigma} \nabla_\mu H_{\nu \rho \sigma} \right)\Big|_{\textup{scalars}} \notag \\[4pt]
 & +\frac{3\lambda_0}{2} \tilde{g}^{ij} \tilde{g}^{k l} \left\{ 2 g^{a b} g^{c d} g^{e f} \partial_{i} \partial_{k} g_{a c} \partial_{j} B_{b e} \partial_{l} B_{d f} \right. \notag \\[4pt]
 & \qquad \qquad \qquad \left. +\frac{1}{2} \partial_{i} g^{a b} \partial_{j} g^{c d} \partial_{k} B_{a c} \partial_{l} B_{b d} + g^{a b} g^{c d} \partial_{i} g^{e f} \partial_{k} g_{ae} \partial_{j} B_{bc} \partial_{l} B_{fd} \right\}.
\end{align}
The remaining terms can only come from a term of the form
\begin{equation}
 R_{\mu \nu \rho \sigma} H^{\mu \nu \eta} {H^{\rho \sigma}}_{\eta},
\end{equation}
which rather miraculously reduces to the precise combination of terms required, i.e.
\begin{align}
 R_{\mu \nu \rho \sigma} H^{\mu \nu \eta} {H^{\rho \sigma}}_{\eta} \longrightarrow -
 \tilde{g}^{ij} \tilde{g}^{k l} \left\{ 2 g^{a b} g^{c d} g^{e f} \partial_{i} \partial_{k} g_{a c} \partial_{j} B_{b e} \partial_{l} B_{d f} +\frac{1}{2} \partial_{i} g^{a b} \partial_{j} g^{c d} \partial_{k} B_{a c} \partial_{l} B_{b d} \right.\ \ & \notag \\[3pt]
 + \left. g^{a b} g^{c d} \partial_{i} g^{e f} \partial_{k} g_{ae} \partial_{j} B_{bc} \partial_{l} B_{fd} \right\}&.
\end{align}
Hence,
\begin{equation}
 L_{\textup{red}_S}\Big|_{\mathcal{O}(\alpha', B^2)} = \lambda_0 \left(-\frac{3}{2} R_{\mu \nu \rho \sigma} H^{\mu \nu \eta} {H^{\rho \sigma}}_{\eta} + \frac{1}{3} \nabla^\mu H^{\nu \rho \sigma} \nabla_\mu H_{\nu \rho \sigma} \right)\Big|_{\textup{scalars}}.
\end{equation}

In summary, we find
\begin{align}
 L_{\textup{red}_S}=&\left( R + 4 g^{\mu \nu} (\partial_{\mu} \phi) (\partial_{\nu} \phi) - \frac{1}{12} H^{2} \right. \notag \\[6pt]
&\quad +\alpha' \lambda_0 \left\{ R^{\mu \nu \rho \sigma} R_{\mu \nu \rho \sigma} 
-\frac{3}{2} R_{\mu \nu \rho \sigma} H^{\mu \nu \eta} {H^{\rho \sigma}}_{\eta} + \frac{1}{3} \nabla^\mu H^{\nu \rho \sigma} \nabla_\mu H_{\nu \rho \sigma} \right. \notag \\[10pt]
&\hspace{20mm} \left. \left.+\frac{1}{8}  H^{\mu \nu \tau} H_{\mu \nu \eta} H^{\rho \sigma \eta} H_{\rho \sigma \tau} + \frac{1}{24}  {H_{\mu}}^{\sigma \tau} {H_{\rho}}^{\mu \nu} {H^{\rho}}_{\tau \eta} {H^{\eta}}_{\nu \sigma} \right\} \right)\Big|_{\textup{scalars}},
\end{align}
which leads us to conclude that the duality complete action up to first order in $\alpha'$ is
\begin{align}
 S=\int \textup{d}x^\mu \sqrt{g} \e^{-2\phi} &\left( R + 4 g^{\mu \nu} (\partial_{\mu} \phi) (\partial_{\nu} \phi) - \frac{1}{12} H^{2} \right. \notag \\[6pt]
&\quad +\alpha' \lambda_0 \left\{ R^{\mu \nu \rho \sigma} R_{\mu \nu \rho \sigma} 
-\frac{3}{2} R_{\mu \nu \rho \sigma} H^{\mu \nu \eta} {H^{\rho \sigma}}_{\eta} + \frac{1}{3} \nabla^\mu H^{\nu \rho \sigma} \nabla_\mu H_{\nu \rho \sigma} \right. \notag \\[10pt]
&\hspace{20mm} \left. \left.+\frac{1}{8}  H^{\mu \nu \tau} H_{\mu \nu \eta} H^{\rho \sigma \eta} H_{\rho \sigma \tau} + \frac{1}{24}  {H_{\mu}}^{\sigma \tau} {H_{\rho}}^{\mu \nu} {H^{\rho}}_{\tau \eta} {H^{\eta}}_{\nu \sigma} \right\} \right). \label{bosR2action1}
\end{align}

In order to compare the action above with the form in which it is usually presented in the literature (see for example \cite{metsaevtseytlin}), consider the following term that appears in the action above:
\begin{equation} \label{simpdhdh1}
 \frac{1}{3} \nabla^\mu H^{\nu \rho \sigma} \nabla_\mu H_{\nu \rho \sigma}.
\end{equation}
Making use of the Bianchi identity
\begin{equation}
\nabla_{[\mu}H_{\nu \rho \sigma]}=0,
\end{equation}
integrating by parts and ignoring boundary contributions throughout gives
\begin{equation} \label{dhdh:eqn1}
 \frac{1}{3} \nabla^\mu H^{\nu \rho \sigma} \nabla_\mu H_{\nu \rho \sigma}= 2 \nabla^{\mu} H^{\nu \rho \sigma} \nabla_{\nu} \phi H_{\mu \rho \sigma} - H^{\nu \rho \sigma} \nabla_{\mu} \nabla_{\nu} {H^{\mu}}_{\rho \sigma}.
\end{equation}
Integrating by parts on the $\mu$ derivative in the first term on the right hand side of the equation above gives
\begin{equation} \label{dhdh:eqn2}
2 \nabla^{\mu} H^{\nu \rho \sigma} \nabla_{\nu} \phi H_{\mu \rho \sigma} = 2 H^{\mu \rho \sigma} {H^{\nu}}_{\rho \sigma} (2 \partial_{\mu} \phi \partial_{\nu} \phi-  \nabla_{\mu}\nabla_{\nu}\phi) - 2 \nabla_{\mu} H^{\mu \rho \sigma} {H_{\rho \sigma}}^{\nu} \partial_{\nu} \phi,
\end{equation}
while anticommuting covariant derivatives in the second term on the right hand side of equation \eqref{dhdh:eqn1} gives
\begin{equation} \label{dhdh:eqn3}
 H^{\nu \rho \sigma} \nabla_{\mu} \nabla_{\nu} {H^{\mu}}_{\rho \sigma} = R_{\mu \nu} H^{\mu \rho \sigma} {H^{\nu}}_{\rho \sigma} - 2 R_{\mu \nu \rho \sigma} H^{\mu \rho \eta} {H^{\nu \sigma}}_{\tau} + \nabla_{\mu} \nabla_{\nu} H^{\nu \rho \sigma} {H^{\mu}}_{\rho \sigma}.
\end{equation}
Substituting equations \eqref{dhdh:eqn2} and \eqref{dhdh:eqn3} into equation \eqref{dhdh:eqn1} gives
\begin{align} 
\frac{1}{3} \nabla^\mu H^{\nu \rho \sigma} \nabla_\mu H_{\nu \rho \sigma}=&R_{\mu \nu \rho \sigma} H^{\mu \nu \eta} {H^{\rho \sigma}}_{\eta} - R_{\mu \nu} H^{\mu \rho \sigma} {H^{\nu}}_{\rho \sigma} + 2 H^{\mu \rho \sigma} {H^{\nu}}_{\rho \sigma} (2 \partial_{\mu} \phi \partial_{\nu} \phi-  \nabla_{\mu}\nabla_{\nu}\phi) \notag \\
&-\nabla_{\mu} \nabla_{\nu} H^{\nu \rho \sigma} {H^{\mu}}_{\rho \sigma} - 2 \nabla_{\mu} H^{\mu \rho \sigma} {H_{\rho \sigma}}^{\nu} \partial_{\nu} \phi, 
\end{align}
where we have used the Bianchi identity satisfied by the Riemann tensor to relate $R_{\mu \nu \rho \sigma} H^{\mu \rho \eta} {H^{\nu \sigma}}_{\tau}$ to $R_{\mu \nu \rho \sigma} H^{\mu \nu \eta} {H^{\rho \sigma}}_{\eta}$.  Finally, using integration by parts on the first term on the second line of the equation above gives
\begin{align}
\frac{1}{3} \nabla^\mu H^{\nu \rho \sigma} \nabla_\mu H_{\nu \rho \sigma}=&R_{\mu \nu \rho \sigma} H^{\mu \nu \eta} {H^{\rho \sigma}}_{\eta} - R_{\mu \nu} H^{\mu \rho \sigma} {H^{\nu}}_{\rho \sigma} + 2 H^{\mu \rho \sigma} {H^{\nu}}_{\rho \sigma} (2 \partial_{\mu} \phi \partial_{\nu} \phi-  \nabla_{\mu}\nabla_{\nu}\phi) \notag \\
& \nabla_{\nu} H^{\nu \rho \sigma} \nabla_{\mu}{H^{\mu}}_{\rho \sigma} - 4 \nabla_{\mu} H^{\mu \rho \sigma} {H_{\rho \sigma}}^{\nu} \partial_{\nu} \phi.  \label{simpdhdh2}
\end{align}
Inserting the equation above into action \eqref{bosR2action1} gives
\begin{align}
 S=\int &\textup{d}x^\mu \sqrt{g} \e^{-2\phi} \left( R + 4 g^{\mu \nu} (\partial_{\mu} \phi) (\partial_{\nu} \phi) - \frac{1}{12} H^{2} \right. \notag \\[6pt]
&+\alpha' \lambda_0 \left\{ R^{\mu \nu \rho \sigma} R_{\mu \nu \rho \sigma} 
-\frac{1}{2} R_{\mu \nu \rho \sigma} H^{\mu \nu \eta} {H^{\rho \sigma}}_{\eta} + \frac{1}{24}  {H_{\mu}}^{\sigma \tau} {H_{\rho}}^{\mu \nu} {H^{\rho}}_{\tau \eta} {H^{\eta}}_{\nu \sigma} -\frac{1}{8} (H^2)^{\mu\nu} (H^2)_{\mu\nu} \right. \notag \\[10pt]
&\quad -\left[ R_{\mu \nu} + 2 \nabla_{\mu} \nabla_{\nu} \phi 
-\textstyle{\frac{1}{4}} (H^2)_{\mu\nu} 
-\textstyle{\frac{1}{2}} g_{\mu\nu} \left(R+4\nabla^2\phi-4(\partial\phi)^2-\textstyle{\frac{1}{12}} H^2\right) \right] (H^2)^{\mu\nu} \notag \\[15pt]
&\quad + \left[\nabla_{\rho} H^{\rho \mu \nu} - 2\partial_{\rho}\phi H^{\rho \mu \nu} \right]
\left[\nabla_{\sigma} {H^{\sigma}}_{\mu \nu} - 2\partial_{\sigma}\phi {H^{\sigma}}_{\mu \nu} \right] \notag \\[10pt]
&\quad \left. \left. -\frac{1}{2} \left[ R+4\nabla^2\phi-4(\partial\phi)^2-\textstyle{\frac{1}{12}} H^2 \right] H^2 \ \right\} \right),
\end{align}
where $(H^2)_{\mu\nu}=H_{\mu\rho\sigma} {H_{\nu}}^{\rho\sigma}$ and $\nabla^2\phi=g^{\mu\nu}\nabla_{\mu}\nabla_{\nu}\phi$. Note that the terms in the square brackets in the action above are the equations of motion for the metric, $B$-field and dilaton, respectively.  Thus, they can be removed using a field redefinition of these respective fields.  In particular, carrying out the following field redefinitions
\begin{align}
 \delta g_{\mu\nu}&=\alpha'\lambda_0 (H^2)_{\mu\nu}, \\
 \delta B_{\mu\nu}&=2\alpha'\lambda_0 \left(\nabla_{\rho} H^{\rho \mu \nu} -2\partial_{\rho}\phi H^{\rho \mu \nu} \right), \\
 \delta \phi &= \frac{1}{4} \alpha'\lambda_0 H^2,
\end{align}
the action reduces to
\begin{align}
 S=\int &\textup{d}x^\mu \sqrt{g} \e^{-2\phi} \left( R + 4 g^{\mu \nu} (\partial_{\mu} \phi) (\partial_{\nu} \phi) - \frac{1}{12} H^{2} \right. \notag \\[6pt]
&+\frac{\alpha'}{4} \left. \left\{ R^{\mu \nu \rho \sigma} R_{\mu \nu \rho \sigma} 
-\frac{1}{2} R_{\mu \nu \rho \sigma} H^{\mu \nu \eta} {H^{\rho \sigma}}_{\eta} + \frac{1}{24}  {H_{\mu}}^{\sigma \tau} {H_{\rho}}^{\mu \nu} {H^{\rho}}_{\tau \eta} {H^{\eta}}_{\nu \sigma} -\frac{1}{8} (H^2)^{\mu\nu} (H^2)_{\mu\nu} \right\} \right),
\end{align}
where we have substituted the value of $\lambda_0$ for the bosonic string, i.e. $\lambda_0=1/4$. This is in precise agreement with the form of the corrections presented in \cite{metsaevtseytlin}.

\section{Conclusions}
In this paper, we have argued that duality symmetry places strong constraints on the form of the unreduced theory. In particular, we have shown that consistency of a theory with duality under reduction can be used to complete known terms in the theory. An important application of this idea is in finding higher derivative matter couplings from known higher derivative corrections. We have illustrated this idea for a simple theory, namely closed bosonic string theory.  We explained in full detail how the Riemann squared term in the tree-level $\alpha'$ correction to the low energy effective action can be completed in a very simple and systematic manner, reproducing, up to field redefinitions, precisely the form that is familiar from the literature.

Given the generic nature of our arguments, this method can be applied to a wide range of theories. Dualities are a pervading feature of many theories, whether string theories or field theories, and if the duality symmetry survives quantum corrections, as is expected in many cases, then the arguments in this paper can be used to efficiently complete known terms.

The example we have considered here is meant to provide a test of our arguments and illustrate the simplicity of our method in full detail. It is our intention to apply this work to more physically interesting theories where higher derivative corrections are important. Of particular interest is higher derivative corrections in heterotic string theory where, for example, it has been argued that there are constraints from cosmology on the higher derivative corrections \cite{maedaohta, maleknejadjabbari, GMQS}. Another application which we will report on elsewhere is the completion of the $\mathcal{R}^4$ term in M-theory \cite{greengutperlevanhove, russotseytlin} that gives rise to the 1-loop $\alpha'^3$ correction in type IIA string theory \cite{ST}, which would determine the RR higher derivative couplings. We stress, however, that even higher derivative corrections that are not directly obtained from M-theory upon reduction can be found in this manner. For example, the tree-level $\alpha'^3$ corrections \cite{
GW, grisaru1, *grisaru2} in type IIA can be found by expressing the duality coset in terms of the massless fields of IIA string theory rather than M-theory fields. Equally, this work can be applied to type IIB string theory where the relation to M-theory is not as direct, but the theory nevertheless exhibits duality symmetry under reduction. 

In the pioneering work of Green and Gutperle \cite{greengutperle}, it is shown that perturbative and non-perturbative quantum corrections to the $\mathcal{R}^4$ correction in type IIB string theory can be encoded in an automorphic function associated with the S-duality group SL$(2, \mathbb{Z})$ . Furthermore, there has been recent progress in understanding the automorphic functions for the U-duality groups \cite{Lambert:2006ny,Green:2010wi,Pioline:2010kb,Green:2010sp,Green:2010kv,Gubay:2011jk,Green:2011vz,Fleig:2012zc}, which similarly encode the quantum corrections to string theories in lower dimensions. In light of this work, an interesting question is whether the automorphic functions can be similarly uplifted. In other words, to what extent and in what way does the fact that the string coupling dependent coefficient of higher derivative corrections reduce to an automorphic function constrain the coefficient of higher derivative terms in ten-dimensional string theory. 
 
Thus far, we have not discussed supersymmetry, which also plays an important role in constraining higher derivative corrections (see for example \cite{greensethi, Basu:2008cf}). Given that these two methods must be compatible with each other, it may be fruitful in studying the link between supersymmetry completion and duality completion. A step in this direction may be possible in the framework of references \cite{dWNsu8, Nso16}, where the connection between supersymmetry and duality is shown to be intimate. Indeed they reformulate eleven-dimensional supergravity in a way that is manifestly locally invariant under the denominator of the duality coset. Within this framework, it may be possible to incorporate fermions without even considering reduction.     

Finally, it would be interesting to understand these results within the context of generalised geometry and double field theory.  A generalised geometric formulation of higher derivative corrections has proved challenging thus far\footnote{See \cite{hohmsiegelzwiebach} for recent work on understanding $\alpha'$ corrections to generalised geometry.}; in particular, within the context of closed bosonic string theory and the aim of defining a generalised curvature tensor that incorporates all the relevant fields.  From the perspective of this work, it may be possible to provide a generalised geometric formulation of higher derivative terms using an $\alpha'$ corrected duality group element (see also \cite{hohmzwiebach1}).  The example of tree-level $\alpha'$ corrections in closed bosonic string theory considered in this paper is apt for such an investigation and we hope to consider this in future work.

\paragraph{Acknowledgements}
We would like to thank Michael Green, Ilarion Melnikov, Hugh Osborn, Chris Pope, Savdeep Sethi, and in particular, Axel Kleinschmidt and Malcolm Perry for useful discussions. It is a pleasure to thank the Cook's Branch nature conservancy for their hospitality.

\newpage
\appendix

\section{Dimensional reduction}
\label{dimred}

Duality in dimensionally reduced string theory and eleven-dimensional supergravity plays a crucial role in this paper. In this appendix we give some formulae that are useful for reducing gravitational terms. In particular we list the spin connection components for the following vielbein ansatz
\begin{equation}
 e_{\mu}{}^{\bar{\mu}} =
 \begin{pmatrix}
  \alpha \tilde{e}_{i}{}^{\bar{i}} & b_{i}{}^{b} \hat{e}_{b}{}^{\bar{a}} \\
  0 & \hat{e}_{a}{}^{\bar{a}}
 \end{pmatrix},
 \label{vielans}
\end{equation}
where $\alpha$ is a scalar that takes into account conformal transformations of $e_{i}{}^{\bar{i}}.$ For example, $\alpha$ may depend on the determinant of $\hat{e}_{a}{}^{\bar{a}}$, or on the dilaton so that we can switch between Einstein and string frames. The unbarred and barred Greek indices in the above expression are $D$-dimensional spacetime and tangent space indices, respectively. Lowercase Latin indices from the start of the alphabet $(a, b, \dots )$ are $d$-dimensional indices, while indices from the middle of the alphabet are $(D-d)$-dimensional indices. The barred versions of these indices indicate tangent space indices in $d$-dimensions and $(D-d)$-dimensions, respectively. The vielbeine $\tilde{e}$ and $\hat{e}$ define metrics
\begin{gather}
 \tilde{g}_{ij} = \tilde{e}_{i}{}^{\bar{i}} \tilde{e}_{j}{}^{\bar{j}} \tilde{\eta}_{\bar{i}\bar{j}}, \\
 \hat{g}_{ab} = \hat{e}_{a}{}^{\bar{a}} \hat{e}_{b}{}^{\bar{b}} \delta_{\bar{a} \bar{b}},
\end{gather}
respectively. The Minkowski metric in $(D-d)$-dimensions, diag$(-1, +1, \dots, +1),$ is denoted $\tilde{\eta},$ while $\delta$ is the Kronecker delta symbol. In our notation 
$$\tilde{e}_{\bar{i}}{}^{i}, \hat{e}_{\bar{a}}{}^{a}$$
are the inverse vielbeine. Therefore, the $i, j, \dots$ indices are raised and lowered with $\tilde{g}^{-1}$ and $\tilde{g},$ and $\bar{i}, \bar{j}, \dots$ are raised and lowered with the Minkowski metric. Similarly for the $a, b, \dots$ and $\bar{a}, \bar{b}, \dots$ indices.     

The spin connection $$\omega_{\mu}{}_{\bar{\nu} \bar{\rho}} = \omega_{\mu}{}_{[\bar{\nu} \bar{\rho}]}$$ is given by Cartan's first equation of structure in the absence of torsion:
\begin{equation}
 \d e^{\bar{\mu}} + \omega^{\bar{\mu}}{}_{\bar{\nu}} \wedge e^{\bar{\nu}} = 0,
\end{equation}
or in components
\begin{equation}
 2 \left( \partial_{[\mu} e_{\nu]}{}^{\bar{\mu}} + \omega_{[\mu|}{}^{\bar{\mu}}{}_{\bar{\nu}} e_{|\nu]}{}^{\bar{\nu}} \right) = 0.
\end{equation}
For the vielbein ansatz in equation \eqref{vielans}, Cartan's first structure equation is solved for components of the spin connection 
\begin{align}
\omega_{i}{}_{\bar{i} \bar{j}} &= \tilde{\omega}_{i}{}_{\bar{i} \bar{j}} + \alpha^{-2} \tilde{e}_{[\bar{i}}{}^{k} \tilde{e}_{\bar{j}]}{}^{j} \left(2 \alpha \tilde{g}_{ik} \partial_{j} \alpha + b_{i}{}_{a} \partial_{[j} b_{k]}{}^{a} - \alpha^{2} b_{j}{}^{a} \partial_{a} \tilde{g}_{ik} -2 \alpha \tilde{g}_{ik} b_{j}{}^{a} \partial_{a} \alpha  - b_{i}{}_{a} b_{j}{}^{b} \partial_{b} b_{k}{}^{a} \right),  \\
\omega_{i}{}_{\bar{a} \bar{i}} &=  -\frac{1}{2} \alpha^{-1} \tilde{e}_{\bar{i}}{}^{j} \hat{e}_{\bar{a}}{}^{a} \left(2  \hat{g}_{a b} \partial_{[i} b_{j]}{}^{b} - b_{i}{}^{b} \partial_{j} \hat{g}_{ab} + \alpha^{2} \partial_{a} \tilde{g}_{ij} + 2 \alpha \tilde{g}_{ij} \partial_{a} \alpha + b_{j}{}^{b} \partial_{b} b_{i a} +  b_{i}{}_{b} \partial_{a} b_{j}{}^{b}  \right),\\
\omega_{i}{}_{ \bar{i} \bar{a}} &= - \omega_{i}{}_{\bar{a} \bar{i}}, \\
\omega_{i}{}_{\bar{a} \bar{b}} &= \hat{e}_{[\bar{a}|}{}^{a} \partial_{i} \hat{e}_{a|\bar{b}]} - \hat{e}_{[\bar{a}}{}^{a} \hat{e}_{\bar{b}]}{}^{b} \partial_{a} b_{i b}, \\
\omega_{a}{}_{\bar{i} \bar{j}} &= - \alpha^{-2} \tilde{e}_{\bar{i}}{}^{i} \tilde{e}_{\bar{j}}{}^{j} \hat{g}_{a b} \partial_{[i} b_{j]}{}^{b} + \tilde{e}_{[\bar{i}|}{}^{i} \partial_{a} \tilde{e}_{i| \bar{j}]} + \alpha^{-2} \tilde{e}_{[\bar{i}}{}^{i} \tilde{e}_{\bar{j}]}{}^{j} \hat{g}_{a b} b_{i}{}^{c} \partial_{c} b_{j}{}^{b}, \\
\omega_{a}{}_{\bar{a} \bar{i}} &= \frac{1}{2} \alpha^{-1} \tilde{e}_{\bar{i}}{}^{i} \hat{e}_{\bar{a}}{}^{b} \left(  \partial_{i} \hat{g}_{a b} -  \hat{g}_{b c} \partial_{a} b_{i}{}^{c} - \hat{g}_{a c} \partial_{b} b_{i}{}^{c}  - b_{i}{}^{c} \partial_{c} \hat{g}_{a b} \right), \\
\omega_{a}{}_{\bar{i} \bar{a} } &= - \omega_{a}{}_{\bar{a} \bar{i}}, \\
\omega_{a}{}_{\bar{a} \bar{b}} &=  \hat{\omega}_{a}{}_{\bar{a} \bar{b}}.
\end{align}
In particular, in this paper, we are interested in toroidal reductions, where there exist $d$ Killing vectors. Letting the Killing vector fields be $\partial_{a},$ the components of the spin connection simplify to 
\begin{align}
\omega_{i}{}_{\bar{i} \bar{j}} &= \tilde{\omega}_{i}{}_{\bar{i} \bar{j}} + \alpha^{-2} \tilde{e}_{[\bar{i}}{}^{k} \tilde{e}_{\bar{j}]}{}^{j} \left(2 \alpha \tilde{g}_{ik} \partial_{j} \alpha + b_{i}{}_{a} \partial_{[j} b_{k]}{}^{a} \right),  \\
\omega_{i}{}_{\bar{a} \bar{i}} &=  -\frac{1}{2} \alpha^{-1} \tilde{e}_{\bar{i}}{}^{j} \hat{e}_{\bar{a}}{}^{a} \left(2  \hat{g}_{a b} \partial_{[i} b_{j]}{}^{b} - b_{i}{}^{b} \partial_{j} \hat{g}_{ab} \right),\\
\omega_{i}{}_{ \bar{i} \bar{a}} &= - \omega_{i}{}_{\bar{a} \bar{i}}, \\
\omega_{i}{}_{\bar{a} \bar{b}} &= \hat{e}_{[\bar{a}|}{}^{a} \partial_{i} \hat{e}_{a|\bar{b}]}, \\
\omega_{a}{}_{\bar{i} \bar{j}} &= - \alpha^{-2} \tilde{e}_{\bar{i}}{}^{i} \tilde{e}_{\bar{j}}{}^{j} \hat{g}_{a b} \partial_{[i} b_{j]}{}^{b}, \\
\omega_{a}{}_{\bar{a} \bar{i}} &= \frac{1}{2} \alpha^{-1} \tilde{e}_{\bar{i}}{}^{i} \hat{e}_{\bar{a}}{}^{b} \partial_{i} \hat{g}_{a b}, \\
\omega_{a}{}_{\bar{i} \bar{a} } &= - \omega_{a}{}_{\bar{a} \bar{i}}, \\
\omega_{a}{}_{\bar{a} \bar{b}} &=  0.
\end{align}

The curvature of the spin connection is given by Cartan's second structure equation,
\begin{equation}
 R^{\bar{\mu}}{}_{\bar{\nu}} = d \omega^{\bar{\mu}}{}_{\bar{\nu}} + \omega^{\bar{\mu}}{}_{\bar{\rho}} \wedge \omega^{\bar{\rho}}{}_{\bar{\nu}}.
\end{equation}
From above, we can write components of the Riemann tensor as follows
\begin{align}
 R^{\mu}{}_{\nu \rho \sigma} &= 2 e_{\bar{\mu}}{}^{\mu} e_{\nu}{}^{\bar{\nu}} \left( \partial_{[\rho} \omega_{\sigma]}{}^{\bar{\mu}}{}_{\bar{\nu}} + \omega_{[\rho|}{}^{\bar{\mu}}{}_{\bar{\rho}} \omega_{|\sigma]}{}^{\bar{\rho}}{}_{\bar{\nu}} \right). \label{riem}
\end{align}

In section \ref{secbos}, we are interested in the scalar sector of the reduced theory. Moreover, the reductions are such that they do not alter the coefficient of the Einstein-Hilbert term, i.e.\ $\alpha=1.$ Since $\tilde{\omega}_{i}{}_{\bar{i} \bar{j}}$ contributes to the Ricci scalar of the reduced theory and $b_{i}{}^{a}$ is a one-form from the perspective of the reduced theory we can ignore these terms. Hence, in such a reduction, the only relevant spin connection components for the scalar sector are 
\begin{align}
 \omega_{i}{}_{\bar{a} \bar{b}} &= \hat{e}_{[\bar{a}|}{}^{a} \partial_{i} \hat{e}_{a|\bar{b}]}, \\
\omega_{a}{}_{\bar{a} \bar{i}} &= \frac{1}{2} \tilde{e}_{\bar{i}}{}^{i} \hat{e}_{\bar{a}}{}^{b} \partial_{i} \hat{g}_{a b}, \\
\omega_{a}{}_{\bar{i} \bar{a} } &= - \omega_{a}{}_{\bar{a} \bar{i}},
\end{align}
or
\begin{align}
 \omega_{i}{}^{\bar{a}}{}_{\bar{b}} &= \frac{1}{2} \left( \hat{e}^{\bar{a}}{}^{a} \partial_{i} \hat{g}_{a b} - 2 \partial_{i} \hat{e}_{b}{}^{\bar{a}} \right) \hat{e}_{\bar{b}}{}^{b}, \\
\omega_{a}{}^{\bar{a}}{}_{\bar{i}} &= \frac{1}{2} \tilde{e}_{\bar{i}}{}^{i} \hat{e}^{\bar{a}}{}^{b} \partial_{i} \hat{g}_{a b}, \\
\omega_{a}{}^{\bar{i}}{}_{\bar{a}} &= - \frac{1}{2} \tilde{e}^{\bar{i}}{}^{i} \hat{e}_{\bar{a}}{}^{b} \partial_{i} \hat{g}_{a b}.
\end{align}
From equation \eqref{riem}, we see that the only components of the Riemann tensor that contribute to the scalar sector of the reduced theory are 
\begin{align}
R'_{abcd} &= -\frac{1}{2} \tilde{g}^{ij} \partial_{i} \hat{g}_{a[c|} \partial_{j} \hat{g}_{b|d]}, \label{riemscal1}\\
 R'_{abij} &= - \frac{1}{2} \hat{g}^{cd} \partial_{[i|} \hat{g}_{ac} \partial_{|j]} \hat{g}_{bd}, \label{riemscal2}\\
 R'_{ij ab} &=  R'_{abij}, \label{riemscal3}\\
R'_{aibj} &= - \frac{1}{2} \partial_{i} \partial_{j} \hat{g}_{ab} + \frac{1}{4} \hat{g}^{cd} \partial_{i} \hat{g}_{bd} \partial_{j} \hat{g}_{ac}, \label{riemscal4} \\
 R'_{aij b} &= - R'_{aibj}, \label{riemscal5}\\
 R'_{iabj} &= - R'_{aibj}, \label{riemscal6} \\
 R'_{iaj b} &= R'_{aibj}. \label{riemscal7}
\end{align}
 The prime denotes the fact that we are only considering the terms in the component that belong to the scalar sector of the reduced theory. 

\pagebreak

\bibliography{highderiv}
\bibliographystyle{utphys}
\end{document}